%% file: paper.tex
\documentclass[twocolumn]{aastex63}
\usepackage{verbatim}
\usepackage{comment} 
\usepackage{rotating}
\include{definitions}

\makeatletter
\def\@hex@@Hex#1%
 {\if a#1A\else \if b#1B\else \if c#1C\else \if d#1D\else
  \if e#1E\else \if f#1F\else #1\fi\fi\fi\fi\fi\fi \@hex@Hex}
\makeatother
\definecolor{afcolor}{HTML}{b3443c}

\definecolor{cpcolor}{HTML}{3cb356}

\received{\today}
\revised{\today}
\accepted{\today}
\submitjournal{MNRAS}

\shorttitle{Dusty REBELS galaxies}
\shortauthors{Ferrara et al.}
\begin{document}

\title{The ALMA REBELS Survey. Epoch of Reionization giants: properties of dusty galaxies at $z \approx 7$}
%\title{Epoch of Reionization giants: properties of dusty REBELS galaxies}

\correspondingauthor{Andrea Ferrara}
\email{andrea.ferrara@sns.it}

% Lead authors
\author[0000-0002-9400-7312]{A. Ferrara}
\affil{Scuola Normale Superiore,  Piazza dei Cavalieri 7, 50126 Pisa, Italy}

\author[0000-0002-2906-2200]{L. Sommovigo}
\affil{Scuola Normale Superiore,  Piazza dei Cavalieri 7, 50126 Pisa, Italy}

\author[0000-0001-8460-1564]{P. Dayal}
\affil{Kapteyn Astronomical Institute, University of Groningen, 9700 AV Groningen, The Netherlands}

\author[0000-0002-7129-5761]{A. Pallottini}
\affil{Scuola Normale Superiore,  Piazza dei Cavalieri 7, 50126 Pisa, Italy}

% Builders providing comments
\author[0000-0002-4989-2471]{R.J. Bouwens}
\affil{Leiden Observatory, Leiden University, NL-2300 RA Leiden, Netherlands}

\author[0000-0002-3120-0510]{V. Gonzalez}
\affil{Departmento de Astronomia, Universidad de Chile, Casilla 36-D, Santiago 7591245, Chile}
\affil{Centro de Astrofisica y Tecnologias Afines (CATA), Camino del Observatorio 1515, Las Condes, Santiago, 7591245, Chile}

\author[0000-0003-4268-0393]{H. Inami}
\affil{Hiroshima Astrophysical Science Center, Hiroshima University, 1-3-1 Kagamiyama, Higashi-Hiroshima, Hiroshima 739-8526, Japan}

\author[0000-0001-8034-7802]{R. Smit}
\affil{Astrophysics Research Institute, Liverpool John Moores University, 146 Brownlow Hill, Liverpool L3 5RF, UK}

% Builders not providing comments
\author[0000-0003-3917-1678]{R.A.A. Bowler}
\affil{Astrophysics, The Denys Wilkinson Building, University of Oxford, Keble Road, Oxford, OX1 3RH, UK}

\author[0000-0003-4564-2771]{R. Endsley}
\affil{Steward Observatory, University of Arizona, 933 N Cherry Ave, Tucson, AZ 85721, USA}

\author[0000-0001-5851-6649]{P. Oesch}
\affil{Observatoire de Gen{\`e}ve, 1290 Versoix, Switzerland}
\affil{Cosmic Dawn Center (DAWN), Niels Bohr Institute, University of Copenhagen, Jagtvej 128, K\o benhavn N, DK-2200, Denmark}

\author[0000-0001-9746-0924]{S. Schouws}
\affil{Leiden Observatory, Leiden University, NL-2300 RA Leiden, Netherlands}

\author{D. Stark}
\affil{Steward Observatory,  University of Arizona, 933 N Cherry Ave, Tucson, AZ 85721, USA}

\author[0000-0001-7768-5309]{M. Stefanon}
\affil{Leiden Observatory, Leiden University, NL-2300 RA Leiden, Netherlands}

% Members providing comments
\author[0000-0002-6290-3198]{M. Aravena}
\affil{Nucleo de Astronomia, Facultad de Ingenieria y Ciencias, Universidad Diego Portales, Av. Ejercito 441, Santiago, Chile}

\author[0000-0001-9759-4797]{E. da Cunha}
\affil{International Centre for Radio Astronomy Research, University of Western Australia, 35 Stirling Hwy, Crawley,26WA 6009, Australia}
\affil{ARC Centre of Excellence for All Sky Astrophysics in 3 Dimensions (ASTRO 3D), Australia}

\author[0000-0001-9419-6355]{I. De Looze}
\affil{Sterrenkundig Observatorium, Ghent University, Krijgslaan 281 - S9, 9000 Gent, Belgium}
\affil{Dept. of  Physics \& Astronomy, University College London, Gower Street, London WC1E 6BT, UK}

\author[0000-0001-7440-8832]{Y. Fudamoto}
\affil{Observatoire de Gen{\`e}ve, 1290 Versoix,  Switzerland}
\affil{Research Institute for Science and Engineering, Waseda University, 3-4-1 Okubo, Shinjuku, Tokyo 169-8555, Japan}
\affil{National Astronomical Observatory of Japan, 2-21-1,  Osawa, Mitaka, Tokyo, Japan}

\author[0000-0002-9231-1505]{L. Graziani}
\affil{Dipartimento di  Fisica, Sapienza, Universit\'a di Roma, Piazzale Aldo Moro 5, I-00185 Roma, Italy}
\affil{INAF/Osservatorio Astrofisico di Arcetri, Largo E. Fermi 5, 50125 Firenze, Italy}

\author[0000-0001-6586-8845]{J. Hodge}
\affil{Leiden Observatory, Leiden University, NL-2300 RA Leiden, Netherlands}

\author[0000-0001-9585-1462]{D. Riechers}
\affil{I. Physikalisches Institut, Universit\"at zu K\"oln, Z\"ulpicher Strasse 77, D-50937 K\"oln, Germany}

\author[0000-0001-9317-2888]{R. Schneider}
\affil{Dipartimento di Fisica, Sapienza, Universit\'a di Roma, Piazzale Aldo Moro 5, I-00185 Roma, Italy}
\affil{Sapienza School for Advanced Studies, Viale Regina Elena 291, 00161 Roma Italy}
\affil{INAF/Osservatorio Astronomico di Roma, via Frascati 33, 00078 Monte Porzio Catone, Roma, Italy}
\affil{Istituto Nazionale di Fisica Nucleare, Sezione di Roma1, Piazzale Aldo Moro 2, 00185 Roma Italy} 

% Members not providing comments
\author[0000-0002-4205-9567]{H.S.B. Algera}
\affil{Hiroshima Astrophysical Science Center, Hiroshima University, 1-3-1 Kagamiyama, Higashi-Hiroshima, Hiroshima 739-8526, Japan}

\author[0000-0003-1641-6185]{L. Barrufet}
\affil{Observatoire de Gen{\`e}ve, 1290 Versoix, Switzerland}

\author[0000-0002-6488-471X]{A.P.S. Hygate}
\affil{Leiden Observatory, Leiden University, NL-2300 RA Leiden, Netherlands}

\author[0000-0002-2057-5376]{I. Labb{\'e}}
\affil{Centre for Astrophysics \& Supercomputing, Swinburne University of Technology, PO Box 218,  Hawthorn, VIC 3112, Australia}

\author[0000-0003-4404-9589]{C. Li}
\affil{Department of Astronomy \& Astrophysics, The  Pennsylvania State University, 525 Davey Lab, University Park, PA  16802, USA}
\affil{Institute for Gravitation and the  Cosmos, The Pennsylvania State University, University Park, PA  16802, USA}

\author[0000-0003-2804-0648]{T. Nanayakkara}
\affil{Centre for Astrophysics \&  Supercomputing, Swinburne University of Technology, PO Box 218,  Hawthorn, VIC 3112, Australia}

\author[0000-0001-8426-1141]{M. Topping}
\affil{Steward Observatory,  University of Arizona, 933 N Cherry Ave, Tucson, AZ 85721, USA}

\author[0000-0001-5434-5942]{P. van der Werf}
\affil{Leiden Observatory, Leiden University, NL-2300 RA Leiden, Netherlands}

\begin{abstract}
We analyse FIR dust continuum measurements for 14 galaxies  (redshift $z\approx 7$) in the ALMA REBELS Large Program to derive their physical properties. Our model uses three input data, i.e. (a) the UV spectral slope, $\beta$, (b) the observed UV continuum flux at $1500$\AA, $\FUV$,  (c) the observed continuum flux at $\approx 158\mu$m, $\FIR$, and considers Milky Way (MW) and SMC extinction curves, along with different dust geometries. We find that REBELS galaxies have $28-90.5$\% of their star formation obscured;  the \textit{total} (UV+IR) star formation rates are in the range $31.5 < {\rm SFR}/(\msunyr) < 129.5$. The sample-averaged dust mass and temperature are $(1.3\pm 1.1)\times 10^7 M_\odot$ and $52 \pm 11$ K, respectively. However, in some galaxies dust is particularly abundant (REBELS-14, $M'_d \approx 3.4 \times 10^7 M_\odot$), or hot (REBELS-18, $T'_d \approx 67$ K). The dust distribution is compact ($<0.3$ kpc for 70\% of the galaxies). The inferred dust yield per supernova is $0.1 \le y_d/M_\odot \le 3.3$, with 70\% of the galaxies requiring $y_d < 0.25 M_\odot$. Three galaxies (REBELS-12, 14, 39) require $y_d > 1 M_\odot$, which is likely inconsistent with pure SN production, and might require dust growth via accretion of heavy elements from the interstellar medium. With the SFR predicted by the model and a MW extinction curve, REBELS galaxies detected in [CII] nicely follow the local $L_{\rm CII}-$SFR relation, and are approximately located on the Kennicutt-Schmidt relation. The sample-averaged gas depletion time is of $0.11\, y_P^{-2}$ Gyr, where $y_P$ is the ratio of the gas-to-stellar distribution radius. For some systems a solution simultaneously matching the observed ($\beta, \FUV, \FIR$) values cannot be found. This occurs when the index $I_m = (F_{158}/\FUV)/(\beta-\beta_{\rm int})$, where $\beta_{\rm int}$ is the intrinsic UV slope, exceeds $I_m^*\approx 1120$ for a MW curve. For these objects we argue that the FIR and UV emitting regions are not co-spatial, questioning the use of the IRX-$\beta$ relation. 
\end{abstract}


\section{Introduction} \label{sec:intro}
%\LS{This intro is BEAUTIFUL} 
Although many aspects of the process, such as e.g. the detailed time evolution, topology of ionized regions, ionizing (LyC) photons emissivity and their escape fraction from early galaxies, are still largely unknown, there is a large consensus \citep[for a discussion, see][]{Ciardi05, Wise19} that reionization was predominantly powered by photons emitted by stars within small galaxies in $10^{8-9} M_\odot$ dark matter halos \citep[][]{Choudhury08, Mitra15, Robertson15}.

Identifying the reionization sources with available instrumentation has represented one of the most intense areas of activity  \citep[reviewed in][]{Dayal18} in observational cosmology in the last two decades. Such searches typically exploit two distinct techniques, fully described in e.g. \citet[][]{Dunlop13}, based on the detection of either (a) the 912\AA\ Lyman break caused by the absorption of LyC photons by interstellar/intergalactic neutral gas\footnote{More precisely, at $z>4$, the Ly$\alpha$ forest becomes the dominant factor, and at $z>6$ we observe a complete Ly$\alpha$ break.}, or (b) the prominent Ly$\alpha$ line emission. These two strategies yield separate \citep[but likely overlapping, see][]{Dayal12} samples of EoR galaxies known as Lyman Break Galaxies (LBG) and Lyman Alpha Emitters (LAE). 

These combined efforts were amazingly successful, and they have allowed to build the UV and Ly$\alpha$ luminosity functions (LF) of EoR galaxies down to very faint magnitudes. For example, at $z \simeq 7$ the UV LF is now sampled over about 8 magnitudes in the range $-22.6 < M_{\rm UV} < -14.5$ \citep[][]{Ono18}. These observations used a combination of space-born surveys using HST \citep{Oesch10,McLure13,Bouwens14}, in some cases exploiting gravitational lensing of massive galaxy clusters, such as the Hubble Frontier Fields  \citep[][]{Lotz17, Livermore17, Ishigaki18, Bhatawdekar21} {  and other similar programs such as RELICS} \citep{Salmon18}, and ground telescopes \citep{Bowler15,Bouwens16}. A full description of the development in the field along with the theoretical implications can be found in \citet[][]{Dayal18}. A comprehensive review of the present knowledge of the LF up to $z=9$ is given in \citet[][]{Bouwens21}; \citet[][]{Oesch18} and \citet[][]{Bowler20} obtained preliminary determinations at even earlier times (up to $z\approx 10$).    

These studies have produced a better characterisation of the faint-end of the LF carrying information on the reionization sources. In addition, they have enabled the  investigation of the EoR \quotes{giants}, i.e. luminous, massive galaxies  \citep[][]{Naidu20, Trebitsch20} shaping the bright end of the LF. {  Using the ULTRAVISTA and VIDEO surveys, \citet[][]{Bowler20} found a steepening of the $z>5$ LF bright-end slope over the $-23 < M_{\rm UV} < -17$ range.}

Indeed, if dust is present already at these very early cosmic epochs, its effects are expected to be more evident on the most massive galaxies. This is because to a first approximation, and for a fixed dust yield, the dust mass is expected to be proportional to the stellar mass. 

The presence of sizeable amounts of dust has a strong impact on galaxy evolution. Dust governs the interstellar medium (ISM) thermal balance \citep{Draine03, Galliano18} by providing photoelectric heating, it controls
important chemical processes such as the formation of H$\ped{2}$ \citep{Tielens10}, which in turn drives molecular chemistry. 
Thus, determining the dust content of EoR galaxies is crucial to interpret observations  \citep{Mancini16, Behrens18, Wilkins18, Arata19, Vogelsberger20, Inoue20, diMascia21, diMascia21b, Shen21}, and build a coherent picture of early galaxy formation. 

Grains absorb the stellar UV light and re-radiate it in the infrared, shielding the dense gas, ultimately triggering the formation of molecular clouds where new stars are born.
However, determining the mass content purely from UV observations is difficult \citep[][]{Calzetti01, Cortese06}. Combining UV and IR observations is then fundamental to constrain the dust content and, even more crucially, the optical and physical properties of the grains, such as their temperature, size distribution and composition \citep{Draine03}. In turn, such information holds the key to investigate the ISM of the first galaxies, and clarify the energy and mass exchange of these systems with the surrounding environment. 

Last but not least, the origin and rapid cosmic evolution of dust is a long-standing question \citep{Todini01, Ginolfi18, Lesniewska19} that has received only partial answers.    

Multi-wavelength observations of galaxies are routinely performed locally \citep[see, e.g.,][]{Cormier19}, and have enabled mighty insights on the physical properties of these systems, and the processes regulating them, particularly for what concerns their ISM. This approach becomes increasingly challenging towards high redshift due to both the faintness of the sources, and the redshifting of diagnostic spectral features to electromagnetic bands out of reach of available instrumentation. With the advent of ALMA, which will soon followed by JWST, the situation has drastically improved. 

In the last few years ALMA observations \citep{Watson15, Laporte17, Hashimoto19, Bakx20, Faisst20, Hodge20, Gruppioni20, Schouws21, Bakx21} have detected dust thermal emission in \quotes{normal} galaxies\footnote{We recall that the presence of dust at $z>6$ was already ascertained from observations of quasar hosts and massive dusty starbursts; for a review see, e.g. \citet{Casey14}} well into the EoR ($z \approx 7$). The copious IR continuum emission from these early galaxies came, at least partly, as a surprise. Naively, a common expectation was that these remote galaxies are low-metallicity, low-dust content systems, although some high-resolution simulations have suggested that enrichment could be very fast, particularly close to star forming regions \citep{Behrens18, Wilkins18, Pallottini19, Graziani20}. 

The dust mass of detected EoR galaxies remains nevertheless uncertain, due to the degeneracy with dust temperature; estimates are in the range $10^{6-8} M_\odot$. Interestingly, a quasi-linear, global redshift evolution of the dust temperature is suggested by some studies \citep{Magdis12, Schreiber18, Bethermin20, Faisst20}. However, the errors in the existing data are large and do not allow to draw firm conclusions yet\footnote{Similar uncertainties hold also for sub-millimeter galaxies, \citet[see, e.g.][]{Dudzeviciute20}}. A statistically significant sample of EoR galaxies is a precondition to assess the existence and physical nature of the relation. 

In this context, the ALMA Reionization Era Bright Emission Line Survey (REBELS) Large Program \citep{bouwens:2021} provides a unique sample of the most massive star-forming galaxies at $z>6.5$. REBELS targets 40 of the brightest (and most robust) galaxies identified over a 7 deg$^2$ area of the sky, and systematically scanning these galaxies for bright ISM-cooling lines (such as the [CII]   158$\mu$m and [OIII] 88$\mu$m) and dust-continuum emission. In the first paper from the collaboration, \citet[][]{Fudamoto21} reported the discovery of two dust-obscured star-forming galaxies at $z = 6.6813 \pm 0.0005$ and $z = 7.3521 \pm 0.0005$. These objects are not detected in existing rest-frame UV data and were discovered only through their [CII] line and dust continuum emission as companions to typical UV-luminous galaxies at the same redshift.

The goal of this paper is to use the newly acquired REBELS data in combination with pre-existing UV data to infer the physical properties of these early systems in a self-consistent manner. Such data are partially already presented in \citet[][]{bouwens:2021}, but more details will be given in Schouws et al. 2022, in prep.; Inami et al. 2022, in prep.; Stefanon et al. 2022, in prep.

In addition to their dust content and temperature, this study also aims at determining the obscured star formation fraction, and constraining the dust yields from the major dust factories in the EoR, i.e. supernovae. In two companion theoretical papers of the REBELS Collaboration, we discuss the implications for the UV luminosity function (Dayal et al. 2022, in prep.), and dust temperature redshift evolution (Sommovigo et al. 2022, in prep).  

The plan of the paper is as follows. In Sec. \ref{sec:Method} we describe the method to derive the relevant physical galaxy quantities from UV and FIR data.
The results are presented in Sec. \ref{sec:Results}, and their additional implications discussed in Sec. \ref{sec:implications}. Sec. \ref{sec:failure} contains a discussion on the use of the IRX-$\beta$ relations and the multi-phase nature of some of the REBELS galaxies. Finally, a brief summary is provided in Sec. \ref{sec:summary}.
For consistency with the data analysis, we adopt the following cosmological parameters: $\Omega_m = 0.3, \Omega_\Lambda=0.7, h=0.7$.

\section{Method}\label{sec:Method}
In this Section we describe the analytical method used to derive the properties of the sample galaxies from the REBELS data. It consists of several steps, each of which is detailed in the following subsections.  

The method uses as an input three quantities (in addition to the source redshift $z_s$) measured by REBELS: (a) the UV spectral slope, $\beta$, such that the specific luminosity  $L_\lambda \propto \lambda^\beta$ in the wavelength range $1600-2500$\AA; (b) the observed UV continuum flux, $\FUV$ at $1500$\AA,  (c) the observed far-infrared continuum flux, $\FIR$, at $\approx 158\mu$m; we express fluxes in units of $\mu$Jy. These data are presently available for 14 galaxies\footnote{These are the only galaxies with dust continuum detection out of the 40 targets of the full REBELS sample. Among the 14 galaxies listed in Tab. \ref{tab:DATA}, 13 have also a [CII] line measurement.  REBELS-06 is undetected in [CII] and therefore only the photo-$z$ is available for this source. However, observations of the source
are still ongoing and [CII] may still be found.\label{foot:REB06}} in the REBELS sample, whose mean redshift is $\langle z_s \rangle = 7.01$.  
A summary of the relevant data for the galaxy in the sample is given in Tab. \ref{tab:DATA}. Although our model does not make predictions on the stellar mass, in the Table we also report the $M_*$ values obtained by the REBELS Collaboration (Stefanon+21, in prep.) from rest-frame UV SED fitting\footnote{The authors adopt a constant star formation history, $Z=0.2 Z_\odot$, a Calzetti et al. (2000) dust extinction law, and a Chabrier (2003) $0.1-300 M_\odot$ IMF. Note that the correction on $M_*$, in principle required for consistency with the metallicity value and Salpeter IMF used here, is well within uncertainties reported in Tab. \ref{tab:DATA}, and thus it does not significantly affect our results. {We warn that using non-parametric prescriptions for the star formation history might result in $M_*$ values on average up to $\approx 3\times$ higher (Topping et al., in prep).}}  using BEAGLE \citep[][]{Chevallard16}. These are only used to check the consistency of the dust SN yield, $y_d$, reported in the penultimate column of Tab. \ref{tab:MW} and Tab. \ref{tab:SMC}.     
%
% MEASURED QUANTITIES WITH ERRORS
%
\begin{table*}
%\begin{minipage}{170mm}
\begin{center}
\caption{Measured REBELS galaxy properties and adopted relative errors. For $F_{1500}$ we have assumed a 20\% error (corresponding to $\approx 0.2$ mag on $M_{\rm UV})$. Galaxy names in the text are abbreviated as REBxx for conciseness.}
\begin{tabular}{lcccccc}
\hline\hline
\multicolumn{7}{c}{\code{REBELS DATA}}\\
\# ID & $z$ & $\beta$ & $F_{1500}$    & $F_{158}$ & $I_m$ & $\log(M_*/M_\odot)$  \\
\hline
      &      &   &$\mu$Jy        & $\mu$Jy   & & \\
\hline
REBELS-05 &6.496&$-1.29^{+0.36}_{-0.44}$    &0.315$\pm 0.063$          & 67.2$\pm 12.7$  &191.1$^{+75}_{-84}$    &$9.16_{-1.00}^{+0.85}$\\
REBELS-06 &6.800&$-1.24^{+0.67}_{-0.35}$    &0.329$\pm 0.066$          & 76.7$\pm 15.3$  &200.2$^{+122}_{-80}$    &$9.50_{-0.79}^{+0.45}$\\
REBELS-08 &6.749&$-2.17^{+0.58}_{-0.58}$    &0.363$\pm 0.073$          &101.4$\pm 19.8$  &1183.0$^{+457}_{-457}$   &$9.02_{-0.68}^{+0.64}$\\
REBELS-12 &7.349&$-1.99^{+0.48}_{-0.76}$    &0.543$\pm 0.109$          & 86.8$\pm 24.3$  &384.3$^{+161}_{-197}$    &$8.94_{-0.70}^{+0.93}$\\
REBELS-14 &7.084&$-2.21^{+0.41}_{-0.47}$    &0.704$\pm 0.141$          & 60.0$\pm 14.7$  &434.9$^{+159}_{-166}$    &$8.73_{-0.70}^{+0.80}$ \\
REBELS-18 &7.675&$-1.34^{+0.19}_{-0.32}$    &0.448$\pm 0.090$          & 52.9$\pm  9.9$  &110.7$^{+34}_{-40}$    &$9.49_{-0.73}^{+0.56}$\\
REBELS-19 &7.369&$-2.33^{+0.45}_{-0.64}$    &0.242$\pm 0.048$          & 71.2$\pm 20.4$  &3869.6$^{+1544}_{-1719}$   &$8.79_{-0.69}^{+0.69}$\\
REBELS-25 &7.306&$-1.85^{+0.56}_{-0.46}$    &0.263$\pm 0.053$          &259.5$\pm 22.2$  &1772.0$^{+660}_{-585}$   &$9.89_{-0.18}^{+0.15}$\\
REBELS-27 &7.090&$-1.79^{+0.42}_{-0.45}$    &0.359$\pm 0.072$          & 50.6$\pm  9.9$  &229.0$^{+84}_{-86}$    &$9.69_{-0.34}^{+0.25}$ \\
REBELS-29 &6.685&$-1.61^{+0.10}_{-0.19}$    &0.547$\pm 0.109$          & 56.1$\pm 12.9$  &128.8$^{+40}_{-42}$     &$9.62_{-0.19}^{+0.19}$\\
REBELS-32 &6.729&$-1.50^{+0.28}_{-0.30}$    &0.313$\pm 0.062$          & 60.4$\pm 17.1$  &213.2$^{+84}_{-85}$    &$9.55_{-0.37}^{+0.35}$\\
REBELS-38 &6.577&$-2.18^{+0.45}_{-0.42}$    &0.404$\pm 0.081$          &163.0$\pm 22.8$  &1786.5$^{+571}_{-555}$   &$9.58_{-1.27}^{+0.74}$\\
REBELS-39 &6.847&$-1.96^{+0.30}_{-0.28}$    &0.798$\pm 0.160$          & 79.7$\pm 16.2$  &224.1$^{+72}_{-71}$    &$8.56_{-0.57}^{+0.57}$\\
REBELS-40 &7.365&$-1.44^{+0.29}_{-0.36}$    &0.302$\pm 0.060$          & 48.3$\pm 12.9$  &165.3$^{+64}_{-69}$    &$9.48_{-0.99}^{+0.45}$\\
\hline
\label{tab:DATA}
\end{tabular}
\end{center}
%\end{minipage}
\end{table*}

Before we proceed it is necessary to define $L_\lambda$, for which we rely on the stellar population synthesis code \code{STARBURST99} \citep{Leitherer99}. We assume: (a) continuous star formation, (b) Salpeter Initial Mass Function in the range $1-100 M_\odot$, (c) metallicity $Z=0.004$; all quantities below are computed at a fixed age of 150 Myr. 
%\LS{two potential issues/caveats: i) very young ages suggested for some of the REBELS galaxies, ii) in all the observational papers they use a different IMF, also for the computation of $M_{\star}$. From slack it also seems that they are assuming a different metallicity, $Z=1/5\ Z_{\odot}$ with $Z_{\odot}=0.2$.} 
The models adopt Geneva stellar tracks including the early AGB evolution up to the first thermal pulse for
masses  $>1.7 M_\odot$. For later use, this IMF produces $\nu_{\rm SN} = (52.89 M_\odot)^{-1}$ supernovae (SN) per unit stellar mass formed.  
% $\nu_{\rm SN} = (134.74 M_\odot)^{-1}$ : value for Salpeter 0.1-100 Msun 

From the specific luminosity we obtain the conversion factor\footnote{The choice of the Salpeter IMF is motivated by consistency with the standard conversion from specific UV luminosity to SFR ${\rm SFR} [M_\odot {\rm yr}^{-1}] = 1.4 \times 10^{-28} L_\nu(1500 {\rm \AA})\ [{\rm erg\, s}^{-1} {\rm Hz}^{-1}]$ typically used by high-$z$ surveys \citep[e.g.][]{Oesch14} which is based on such IMF \citep{Kennicutt98, Madau14}. Written in this form, the numerical factor corresponding to our ${\cal K}_{1500}$ is equivalent to $4.45\times 10^{-29}$; the $\approx 3\times$ difference arises from the subsolar ($Z=0.004$) metallicity assumed here.} between the intrinsic (i.e. unattenuated) luminosity, $L^*_{1500}$, and the star formation rate (SFR) at $\lambda_{1500} =c/\nu_{1500}= 1500$\AA\,:
\be\label{eq:K1500}
{\cal K}_{1500} \equiv \frac{L^*_{1500}}{\rm SFR} =\frac{(\lambda L^*_\lambda)_{\lambda_{1500}}}{\rm SFR} = 1.174 \times 10^{10}; 
\ee
${\cal K}_{1500}$ has units of ${L_\odot}/(M_\odot {\rm yr}^{-1})$.
%for comparison with other works, the value in eq. \ref{eq:K1500} is equivalent to $2.4\times 10^{28} {\rm erg\, s}^{-1} {\rm Hz}^{-1}/(M_\odot {\rm yr}^{-1})$ at 1500\AA.  
{  We have tested the dependence of ${\cal K}_{1500}$ on the star formation history by considering a ${\rm SFR}(t)\propto \exp(t/t_0)$. This form allows to model both increasing (when the timescale $t_0 >0$) and decreasing ($t_0<0$) histories; we have explored the range $1\, {\rm Myr}\le \vert t_0\vert < 100\, {\rm Myr}$. Increasing histories yield up to 30\% higher ${\cal K}_{1500}$ values with respect to the constant SFR one; a decreasing SFR lowers ${\cal K}_{1500}$ by a factor $\simlt 2$. As high-$z$ galaxies generally feature a time-increasing SFR \citep{Pallottini19, Pallottini22}, we consider the value in eq. \ref{eq:K1500} a reasonable choice.}

Similarly, we compute the intrinsic UV spectral slope, 
\begin{equation}\label{eq:beta_int}
\beta_{\rm int} \equiv \frac{d\ln\,L_{\lambda}}{d\ln\,\lambda} =\frac{\ln(L^*_{\lambda_1}/L^*_{\lambda_2})}{\ln({\lambda_1}/{\lambda_2})} = -2.406,
\end{equation}
where $(\lambda_1, \lambda_2) = (1600, 2500)$\AA.
From the input data we build a non-dimensional \textit{molecular index}, 
\begin{equation}\label{eq:Im}
    I_{\rm m} = \frac{(F_{158}/F_{1500})}{(\beta-\beta_{\rm int})},
\end{equation}
whose physical meaning will be discussed later on. The derived values of $I_{\rm m}$ are given in Tab. \ref{tab:DATA} for each galaxy in the sample. 
% ______________________________________________________
\subsection{From UV slope to optical depth}\label{subsec:beta-to-tau}
Converting $\beta$ into a dust optical depth, $\tau_\lambda$, involves the knowledge of the dust extinction curve, and a model for the radiative transfer (RT) of stellar radiation through dust. We adopt two extinction curves, both taken from \citet[WD01]{Weingartner01}, appropriate for the (a) Milky Way with extinction factor $R_V=3.1$ (hereafter MW), and (b) SMC bar (SMC). 

The extinction cross-sections are normalized to the SB99 metallicity $Z=0.004$ by assuming a linear scaling with metallicity for which we set $Z_{\rm MW}=0.0142$ \citep[i.e., the solar value,][]{Asplund09} and $Z_{\rm SMC}=1.63\times 10^{-3}$ \citep{Choudhury18} for the MW and SMC, respectively.  

By generalizing eq. \ref{eq:beta_int} to the observed (attenuated) luminosity, $L_\lambda = L^*_\lambda T_\lambda$, where $T_\lambda(\tau_\lambda)$ is the UV dust transmissivity (see below) one can write  
\be\label{eq:beta}
\beta = \beta_{\rm int} - 2.24  \ln \left(\frac{T_{\lambda_1}}{T_{\lambda_2}}\right);
\ee
note that for a flat transmissivity ($T_\lambda \approx$ const.), $\beta = \beta_{\rm int}$, independently of optical depth.

We pause to emphasise that there is a distinction between the \textit{physical} value, $\tau_\lambda$, of the optical depth (entering eqs. \ref{eq:slab} and \ref{eq:screen} below), and the \textit{effective} one deduced from the flux attenuation, $\tau_{\rm eff} = - \ln T$. While $\tau_{\rm eff}$ depends on the radiative transfer properties, $\tau_\lambda$ is determined, as we will see below, by the galaxy dust mass and distribution only. 

The functional form of $T_\lambda$ depends on radiative transfer, and therefore on the optical properties of dust grains (the wavelength-dependent albedo, $\omega$, and asymmetry parameter, $g$, i.e. the mean cosine of the scattering angle), and relative spatial distribution of stars and dust\footnote{The values of $\omega$ and $g$ are consistently computed from the adopted WD01 MW and SMC curves. For reference, at 1500\AA: $\omega = (0.3807, 0.4170), g = (0.6633,0.5756)$ for (MW, SMC).}. To bracket such uncertainty we consider two possibilities (although we have experimented with additional ones): (a) a slab geometry in which stars and dust are mixed but have different scale heights; (b) a spherical dust distribution with a central source including scattering. 

These two geometries should mimic the two typical evolutionary stages, both predicted \citep{Pallottini19, Kohandel19, Kohandel20} and observed \citep{Jones17, Smit18} in high-$z$ galaxies, resulting from a periodic switch from a well-designed proto-disk (slab geometry), and a more isotropic (spherical) configuration induced by frequent merging events.

$\square$ \textit{Slab geometry.} \citet{Baes01} found the solution for a slab of total optical depth $\tau_\lambda$, in which dust and stars both follow a vertical exponential distribution with a ratio of scale heights $\zeta = h_d/h_*$. The transmissivity is 
\begin{equation}\label{eq:slab}
T_\lambda = \frac{1}{\mu}\,{e^{-\frac{(1-\omega)}{2\mu}\tau_\lambda}}\,{\cal W}_\zeta\Big[\frac{(1-\omega)\tau_\lambda}{2\mu}\Big].
\end{equation}
In the previous equation, $\mu = \cos \theta$, denotes the direction making an
angle $\arccos \mu$ with the face-on direction $\mu=1$. We fix this value to the most probable inclination for randomly oriented galaxies, $\mu = \cos(\pi/2-1)= 0.841$. Finally\footnote{Note that if stars and dust are homogeneously mixed ($\zeta=1$), and scattering is neglected ($\omega=0$), $W_\zeta = x^{-1}\sinh x$ and eq. \ref{eq:slab} reduces to the more familiar formula
\begin{equation}\label{eq:slab1}
T_\lambda = \frac{(1-e^{-\tau_\lambda/\mu})}{\tau_\lambda}.
\end{equation}},
\begin{equation}\label{eq:W}
W_\zeta(x) = \zeta \int_0^1 (1-t)^{\zeta-1} \cosh(x t) dt.
\end{equation}

$\square$ \textit{Spherical  geometry.} The classical \citet{Code73} solution for {a spherical dust distribution}, obtained with the two-stream approximation and confirmed by Monte Carlo radiative transfer simulations \citep{Ferrara99,Krugel09, diMascia21}, is 
\begin{equation}\label{eq:screen}
    T_\lambda = \frac{2}{(1+\eta)e^{\xi \tau_\lambda} + (1-\eta)e^{-\xi \tau_\lambda}},
\end{equation}
where 
\begin{equation}
\eta     = \sqrt{(1-\omega)/(1-\omega g)};\quad \xi       = \sqrt{(1-\omega)(1-\omega g)}. 
\end{equation}
Eq. \ref{eq:screen} reduces to the standard $e^{-\tau_\lambda}$ screen solution for pure absorption ($\omega=0$).

\begin{figure}
\centering\includegraphics[scale=0.4]{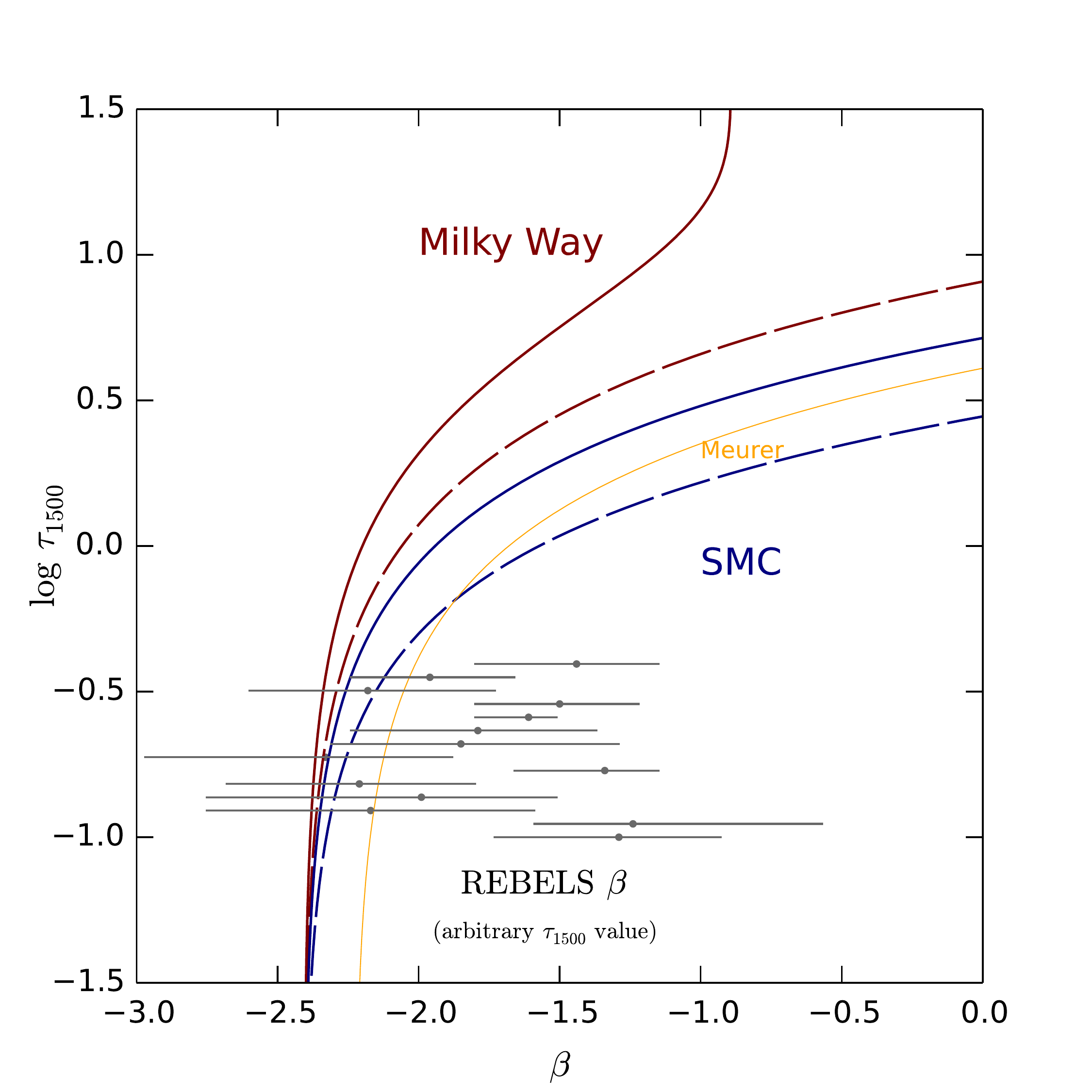}
\caption{Relation between optical depth at 1500\AA, $\tau_{1500}$, and UV spectral slope, $\beta$, often referred to as the \textit{attenuation} curve, for the Milky Way (red curves) and SMC (blue) for a slab  (solid) and spherical (dashed) geometry with $\zeta=2$ (i.e., the dust scale height is twice the stellar one). Also shown for reference is the empirical \citet{Meurer99} relation (orange) $\tau_{1600} = 4.43+1.99\beta$. The $\beta$ values for the 14 REBELS sources with   158$\mum$ continuum detections are shown by grey points/lines at arbitrary $\tau_{1500}$ values.}
\label{Fig:01}
\end{figure}

Armed with the expressions for $T_\lambda$ for the slab and spherical cases, and with the help of eq. \ref{eq:beta}, we can derive the optical depth at 1500\AA, $\tau_{1500}$, of a galaxy with spectral slope $\beta$ (Fig. \ref{Fig:01}). For a given $\beta$ value, slab configurations (we show the case $\zeta=2$ as an example) are more transparent than spherical ones, i.e. they require a larger optical depth to produce the same slope. Also, SMC curves are more opaque than MW ones, i.e. they require less dust to produce the same $\beta$ value. 

Due to their larger transparency, slab geometries cannot produce arbitrarily large $\beta$ values for the MW curve, as indicated by the vertical asymptote at $\beta_{\rm max}\approx -0.9$. The value of $\beta_{\rm max}$ increases with $\zeta$, i.e. as the solution progressively approximates a more opaque screen geometry. Reproducing the measured $\beta$ values in REBELS (grey lines in Fig. \ref{Fig:01}) imposes a lower limit $\zeta \simgt 1.7$ when using the MW curves. For this reason, and because increasing $\zeta$ to values $> 2$ has no effect on our results, we set $\zeta=2$ from now on.  As an illustration, for a MW dust slab the galaxies in the sample have $0.92 < \tau_{1500} < 8.44$, but $1.01 < \tau_{\rm eff} < 2.35$, i.e. they are effectively only mildly optically thin\footnote{The 1500\AA\, to V-band conversion is $\tau_{1500} = (2.655, 5.319) \tau_V$ for (MW, SMC) curves, respectively.}. 

%
% FIGURE 1
%

% ______________________________________________________
\subsection{From optical depth to dust mass}
To derive the dust mass, $M_d$, we start by writing the following relation:
\be\label{eq:dust_mass1}
\tau_{1500} = \frac{1}{\pi m_p f_\mu}\frac{M_d}{r_d^2} \sigma_{1500}, 
\ee
where $m_p$ is the proton mass\footnote{For simplicity, we have assumed a pure hydrogen gas with unity mean molecular weight.}, $r_d$ is the (unknown) radius of the spherical or disk-like (for the slab approximation) dust distribution,  and $\sigma_{1500} = (3.66, 7.25)\times 10^{-22} {\rm cm}^{2}$ is the WD01 (MW, SMC) dust extinction cross section at 1500\AA, scaled to the appropriate metallicity (see Sec. \ref{subsec:beta-to-tau}). 
The factor $f_\mu$ is a geometrical correction; it is $f_\mu=(4/3, \mu)$ for a homogeneous sphere or a disk\footnote{For a disk, this $f_\mu$ expression is valid as long as the viewing angle $\theta < \pi/2-\arctan(h_d/r_d)$, which for our choice of $\mu$ is always satisfied if $h_d/r_d<1$.}, respectively. Inserting the numerical factors, eq. \ref{eq:dust_mass1} is rewritten as 
\be\label{eq:dust_mass2}
M_d = f_\mu \frac{\tau_{1500}}{\tau_0} \left(\frac{r_d}{{\rm kpc}}\right)^2 M_\odot, 
\ee
where $\tau_0 = (1.09, 2.17)\times 10^{-8}$ for (MW, SMC).

% ______________________________________________________
\subsection{Star formation rate and infrared luminosity}
Having determined $\tau_{1500}$ and the corresponding value of $T_{1500}$ from eqs. \ref{eq:slab} and \ref{eq:screen}, for each REBELS galaxy we can straightforwardly derive two important quantities: the total SFR, and the total FIR luminosity, $L_{\rm IR}$, typically computed in the range $8-1000\, \mu$m. Recalling eq. \ref{eq:K1500}, the SFR is obtained from the observed UV flux: 
\be\label{eq:SFR_obs}
F_{1500} = \frac{g(z_s)}{4\pi \nu_{1500}} {\cal K}_{1500}\, T_{1500}\, {\rm SFR};
\ee
$g(z_s) = (1+z_s)/d_L^2$, where $d_L$ is the luminosity distance to the galaxy located at redshift $z_s$.
The observed total FIR luminosity is then
\be\label{eq:L_IR}
L_{\rm IR} =  \alpha \frac{(1-T_{1500})}{T_{1500}} L_{1500} = {\rm IRX}\, L_{1500},
\ee
which shows that the so-called infrared excess parameter ${\rm IRX} \equiv L_{\rm IR}/L_{1500} = \alpha (1-T_{1500})/T_{1500} $. The ${\cal O}(1)$ coefficient $\alpha$ represents the UV bolometric absorption correction (dust is heated also by UV photons with  $\lambda \neq \lambda_{1500})$; for simplicity we set $\alpha=1$ in the following. 

Finally, a quantity that is often reported in the literature is the \textit{observed} star formation fraction, $f_{\rm obs}$. Using the previous results, it is easy to show that $f_{\rm obs}$ can be related to IRX or $T_{1500}$:
\begin{equation}\label{eq:fobs}
    f_{\rm obs} = \frac{L_{1500}}{L_{1500}+ L_{\rm IR}} = \frac{1}{1+ {\rm IRX}} = T_{1500}.
\end{equation}

% ______________________________________________________
\subsection{Dust temperature}
Dust grains are heated by absorption of UV photons and re-emit such energy in the FIR. The emitted radiation spectrum is usually modelled as a grey-body from which the mean dust temperature, $\bar T_d$, can be derived: 
\begin{equation}\label{eq:Tmpt}
    \bar T_d = \left(\frac{L_{\rm IR}}{\Theta M_d}\right)^{1/(4+\beta_d)};
\end{equation}
where
\begin{equation}
    \Theta = \frac{8\pi}{c^2}\frac{\kappa_{158}}{\nu_{158}^{\beta_d}}\frac{k_B^{4+\beta_d}}{h_P^{3+\beta_d}}\zeta(4+\beta_d)\Gamma(4+\beta_d).
\end{equation}
The mass absorption coefficient, $\kappa_\nu = \kappa_{158}(\nu/\nu_{158})^{\beta_d}$ is pivoted at wavelength $\lambda_{158} = c/\nu_{158} =158\mu$m since high-$z$ ALMA observations are often tuned to the rest wavelength of [CII] emission. We take $\kappa_{158}$ and $\beta_d$ consistently with the adopted WD01 extinction curve 
%and metallicity: $\kappa_{158}=(2.93,33.19)\, {\rm cm}^2 {\rm g}^{-1}$, 
$\kappa_{158}=(10.41,13.55)\, {\rm cm}^2 {\rm g}^{-1}$, 
and $\beta_d=(2.03, 2.07)$ 
for (MW, SMC); $\zeta$ and $\Gamma$ are the Zeta and Gamma functions, respectively. Other symbols have the usual meaning. We then find $\Theta = (4.89, 5.33)\times 10^{-6}$ for (MW, SMC). 

We define the temperature in eq. \ref{eq:Tmpt} as the mean \textit{physical} dust temperature. Such value corresponds to the temperature dust grains would attain should the available UV energy being uniformly distributed among them. This is possible only if the system is optically thin, i.e. $\tau_{1500} \ll 1$. In general, though, radiative transfer effects produce a temperature distribution, with $T_d$ decreasing away from the source. Although radiative transfer is a complex problem which can be fully treated with detailed numerical simulations\footnote{{  These two works use the SKIRT code (\url{skirt.ugent.be}) to post-process the simulation outputs.}} \citep[see, e.g.][]{Behrens18, Liang19}, we nevertheless try to approximately take into account this effect for the simple geometry adopted here. 

In App. \ref{app_a} we show that the luminosity-weighted temperature,$\langle T_d \rangle_L$, of an absorbing dust layer depends on its total optical depth, and can be written as
\be\label{eq:TL}
\langle T_d \rangle_L = \bar T_d\, \frac{6}{7}\tau_{1500}^{1/6}\frac{(1 - e^{-7\tau_{1500}/6})}{(1 - e^{\tau_{1500}})^{7/6}}  \equiv \bar T_d f_L(\tau_{1500}).
\ee
For the $\tau_{1500}$ values deduced for REBELS galaxies, applying eq. \ref{eq:TL} results in temperatures that are $\simlt 20$\% higher than $\bar T_d$ (see Fig. \ref{Fig:App_Fig01}). 
As discussed in App. \ref{app_a}, to conserve energy $M_d$ in eq. \ref{eq:dust_mass2} must be reduced by a factor $f_L^{-(4+\beta_d)}$, which might result in a $(2-3)\times$ lower mass estimate. In the following we denote this reduced dust mass by $M_d' = f_L^{{-(4+\beta_d)}} M_d $. 

% ______________________________________________________
\subsection{Observed flux at  158$\mu$m}
From the previous results it is straightforward to compute the rest frame   158$\mu$m specific flux observed at $\lambda = 158(1+z_s)\, \mu$m:
\begin{equation}\label{eq:F158}
    F_{158} = g(z_s)\, \kappa_{158} M'_d [B_{158}(T'_d)-B_{158}(T_{\rm CMB})];
\end{equation}
$B_\lambda$ is the black-body spectrum, and $T_{\rm CMB}(z)=T_0(1+z)$, with $T_0=2.7255$ K  \citep{Fixsen09} is the CMB temperature at redshift $z$. Equation \ref{eq:F158} accounts for the fact that the CMB acts as a thermal bath for dust grains, setting a lower limit to their temperature. At $z=7$ such minimum temperature corresponds to $T_{\rm CMB}=21.8$ K. Finally, $T'_d$ is the CMB-corrected dust temperature\footnote{In the remainder of the paper we will always refer to dust temperature as the \textit{CMB-corrected} one, i.e. $T'_d$ in eq. \ref{eq:CMB}.} following \cite{daCunha13}, 
\begin{equation}\label{eq:CMB}
T'_d = \{\langle T_d \rangle_L^{4+\beta_d}+T_0^{4+\beta_d}[(1+z)^{4+\beta_d}-1]\}^{1/(4+\beta_d)}.
\end{equation}
Note that both $M'_d$ and $T'_d$ depend on the radial extent of the dust distribution, $r_d$ (see eq. \ref{eq:dust_mass2} and \ref{eq:Tmpt}). The latter can be obtained by imposing that $F_{158}$ from eq. \ref{eq:F158} matches the corresponding observed flux for each galaxy in the REBELS sample. 

% __________________________________________________________________
\section{Results}\label{sec:Results}

\begin{figure}
\centering\includegraphics[scale=0.4]{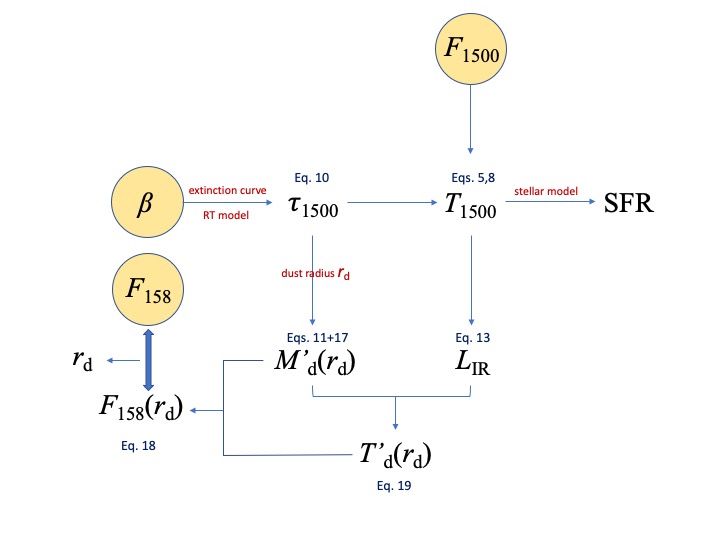}
\caption{Schematic overview of the method. Yellow circles denote the three input observables measured by REBELS; red text indicates assumptions made in that step. Equations defining the different quantities in the text are indicated (blue).}
\label{Fig:02}
\end{figure}

Before discussing the results, let us briefly summarize our method, also illustrated schematically in Fig. \ref{Fig:02}. We use three observables measured by REBELS for 14 galaxies: these are $\beta$, $F_{1500}$, and $F_{158}$. From $\beta$, for given a RT model (i.e. transmissivity $T_{1500}$, also equal to $f_{\rm obs}$) and extinction curve, we determine $\tau_{1500}$, and hence the dust mass $M'_d$, modulo the dust distribution radius, $r_d$. We then use $F_{1500}$ to determine the total SFR and $L_{\rm IR} = {\rm IRX}\, L_{1500}$, which then form the basis to compute the (RT+CMB)-corrected, luminosity-weighted dust temperature $T'_d$, and $F_{158}$. Finally, by imposing that $F_{158}$ matches the observed  158$\mu$m flux, we determine $r_d$ (the only free parameter\footnote{Note that the spatial resolution of the REBELS survey ($\approx 7$ kpc) is much larger than the $r_d \simlt 1$ kpc values obtained here.} of the model once the extinction curve and the RT geometry have been fixed) for each galaxy. 

Hence, from 3 data inputs ($\beta$, $F_{1500}, F_{158}$), our model can predict 7 physical properties for each galaxy ($\tau_{1500},\, T_{1500},\, {\rm SFR},\, L_{\rm IR},\, T'_d,\, M'_d,\, r_d$). By construction, a galaxy with the set of derived properties matches the observed 1500\AA\ and   158$\mu$m continuum data. 

\begin{figure}
\centering\includegraphics[scale=0.37]{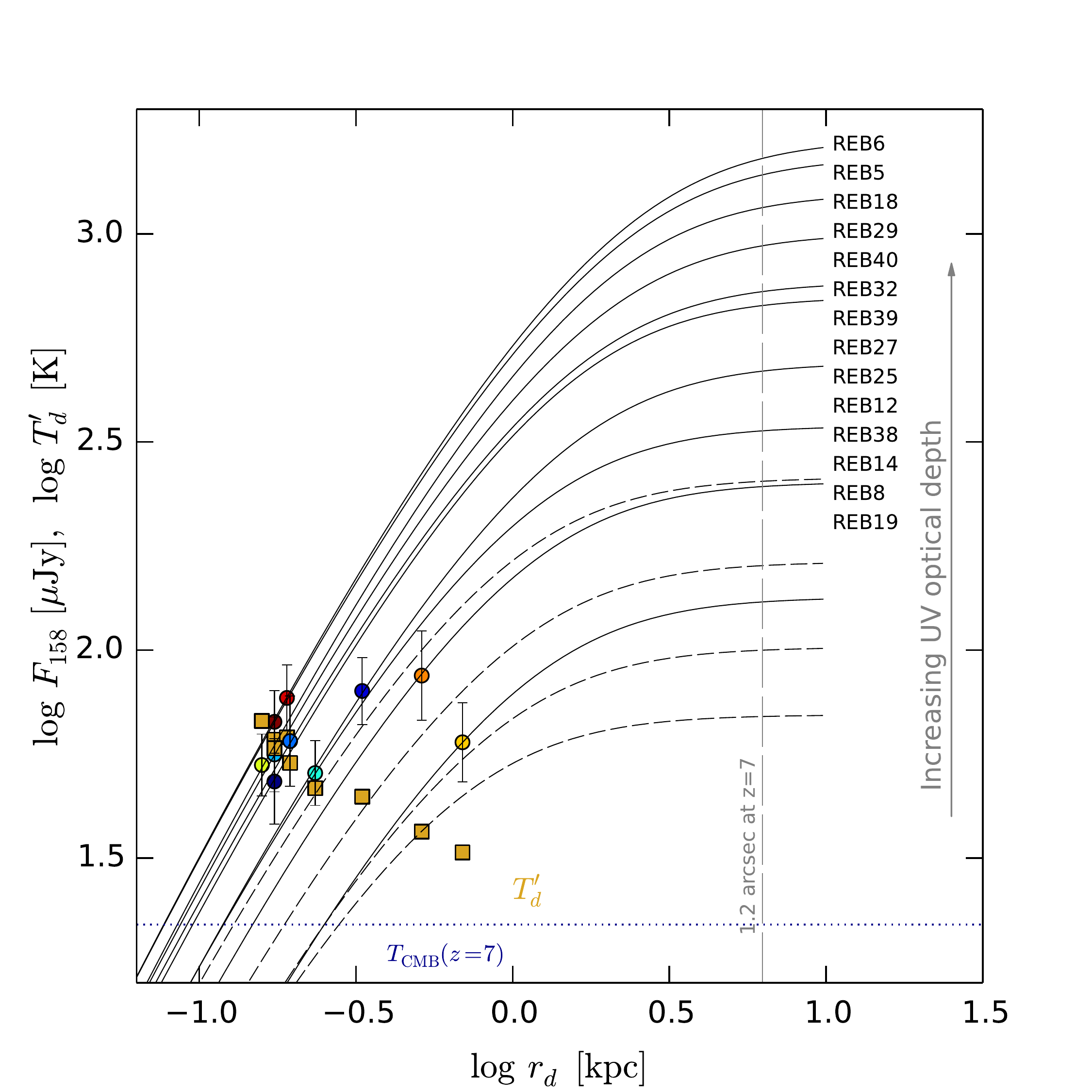}
\caption{Predicted $F_{158}$ flux (curves; ordered as the galaxy name list) as a function of dust radius, $r_d$, for each galaxy in the REBELS sample. Curves  are shown for MW extinction, slab geometry case reported in Tab. \ref{tab:MW}.  The measured value of $F_{158}$ (points) is used to determine $r_d$ for each system (see Fig. \ref{Fig:02}). For 4 galaxies (REBELS-08,19,25,38) no solution can be found as their $F_{158}$ flux is too high to be matched by the corresponding curve (dashed lines). Also shown is the luminosity-weighted dust temperature, $T'_d$ (squares), and the angular resolution (vertical line) of the survey.}
\label{Fig:03}
\end{figure}

Finally, we note that, in some cases, a solution cannot be found because the observed  158$\mu$m flux cannot be retrieved from the input $F_{1500}$ and $\beta$ values. To see this let us inspect Fig. \ref{Fig:03}. As an example, there we consider the MW extinction, slab geometry case. The Figure illustrates the final step of the method sketched in Fig. \ref{Fig:02}, i.e. the determination of $r_d$. This is derived by matching the predicted $F_{158}$ to the observed one for each galaxy. The $F_{158}$ trend with $r_d$ can be understood as follow. As $M'_d \propto r_d^2$, $F_{158}$ initially increases due the larger amount of emitting material. However, $T'_d$ (golden  squares) decreases as the dust distribution becomes more extended until it approaches $T_{\rm CMB}$. At that point $F_{158}$ reaches a plateau independent of $r_d$. The plateau level increases with $\tau_{1500}$: from Fig. \ref{Fig:03} and Tab. \ref{tab:MW} we see that REBELS-06 -- the most opaque ($\tau_{1500}=8.44$) system -- has the potentially highest   158$\mu$m flux, $\approx 3$ mJy.  

Thus, for a given $\tau_{1500}$, the   158$\mu$m flux cannot be arbitrarily large. If the observed $F_{158}$ for a galaxy exceeds this value, the method does not yield a solution. In the MW case, for example, this occurs for 4 (REBELS-08,19,25,38) out of 14 galaxies in the sample (see Tab. \ref{tab:MW}). We discuss the interpretation of these no-solution cases in Sec. \ref{sec:failure}. 
For the well-behaved galaxies we find sub-kpc $r_d$ values implying that at the REBELS spatial resolution ($\approx 1.2$ arcsec or 6.3 kpc at $z=7$) these objects are unresolved. 
%
% FIGURE 2
%

We next discuss our results separately for MW and SMC extinction curves, and highlight the differences induced by the adopted RT (slab/spherical) model for each curve.

%
% FIGURE 3
%

\subsection{Milky Way extinction curve}
The full results for this case are reported in Tab. \ref{tab:MW}, and displayed graphically in Fig. \ref{Fig:04}. First of all we note that our method provides a self-consistent solution matching the data for 10 out of 14 galaxies in the sample. These solutions are presented in the following. The interpretation and implications of no-solution cases will be instead discussed in Sec. \ref{sec:failure}.  
\subsubsection{Slab geometry}
Starting from the slab geometry case, the results indicate that $\tau_{1500}=2.655\tau_V$ varies from 0.92 to 8.44, i.e. galaxy are physically optically thick at 1500\AA\ but mildly so in the V-band and in terms of the effective optical depth, $\tau_{\rm eff}$ (see discussion in Sec. \ref{subsec:beta-to-tau}).  This range translates in transmissivity values going from 9.5\% to 72\%, the most star-forming galaxies being the most obscured. Since $T_{1500} = f_{\rm obs}$ (see eq. \ref{eq:fobs}), we conclude that $28-90.5$\% of the star formation at high-$z$ is obscured, if the REBELS targets fairly sample early galaxy populations.      

The deduced total star formation rates are relatively sustained, $31.5 < {\rm SFR}/(\msunyr) < 129.5$ with REBELS-14 (REBELS-18) being the least (most) star forming system, with a mean ${\rm SFR} = 65.8\, \msunyr$. The ratio between the total IR luminosity and the SFR, which we define as ${\cal K}_{\rm IR}$ paralleling eq. \ref{eq:K1500}, varies from  ${\cal K}_{\rm IR} \simeq 10^{10} L_\odot/(\msunyr)$ for the most star-forming galaxy to ${\cal K}_{\rm IR} \simeq 3.3\times 10^{9} L_\odot/(\msunyr)$ towards the lower end of the SFR range, corresponding to a factor $\approx 3$ decrease. Thus, the specific IR luminosity of low-SFR galaxies is lower than expected for vigorously star-forming ones. We note in passing that the value ${\cal K}_{\rm IR} \simeq 10^{10} L_\odot/(\msunyr)$ matches the commonly assumed one prescribed by \citet[][]{Kennicutt98}.

Our model also constrains dust properties. The 10 systems for which a solution can be found all contain considerable amounts of dust at luke-warm temperatures. The sample-averaged dust mass and temperature are $(1.3\pm 1.1)\times 10^7 M_\odot$ and $52 \pm 11$ K, respectively. However, in some galaxies dust is particularly abundant (REBELS-14, $M'_d \approx 3.4 \times 10^7 M_\odot$), or hot (REBELS-18, $T'_d \approx 67$ K). The dust distribution is compact as one can infer from the values of $r_d$. These are all sub-kpc, with 70\% of the galaxies showing $r_d < 0.3$ kpc. Such compact configuration, in general, produces high dust temperatures: the hottest REBELS-18 indeed is the most compact one with only $5\times 10^6 M_\odot$ of dust concentrated in a very small region ($r_d = 0.16$ kpc). 
As a caveat, we note that the implied compactness might partly be due to the adopted geometry, such that actual observed sizes could be larger if the dust/starlight are more inhomogenously distributed. 

Finally, by using the stellar mass, $M_*$, obtained by the REBELS Collaboration from SED fitting, we estimate the dust yield per supernova\footnote{We neglect dust production from AGB stars ($2 < M/M_\odot < 8$ as their evolutionary time is longer than the Hubble time at $z\approx 7$. Critical issues related to dust growth at early times are discussed in  \citet[][]{Ferrara16, Ferrara21}.}, $y_d = M'_d/\nu_{\rm SN} M_*$. Although \textit{this is not a self-consistent output}, as it uses an independent estimate of $M_*$, it nevertheless represents a useful reference. We find $0.1 \le y_d/M_\odot \le 3.3$ with a mean value $\langle y_d \rangle = 0.9 M_\odot$ with 7 (out of 10) galaxies requiring $y_d < 0.25 M_\odot$. {As already mentioned in Sec. \ref{sec:Method}, stellar masses determined assuming a non-parametric star formation history  might be on average a factor $\approx 3$ larger. In this case the SN yield would be correspondingly decreased to $0.03 \le y_d/M_\odot \le 1.1$.}
Such dust yield
%\footnote{The yield also includes sub-dominant contributions to dust production such as AGB stars and grain growth in the ISM.} 
range is in good agreement with available estimates \citep{Todini01, Hirashita15, Marassi15, Marassi19, Gall18, Lesniewska19} which typically indicate $y_d = 0.01-0.45 M_\odot$ \citep[for a review see][]{Cherchneff14}. Note, however, that even higher values are found, as in the case of the Cas A remnant for which $y_d = 0.99^{+0.10}_{-0.09} M_\odot$ \citep{Priestley19, Niculescu21}, or G54.1+0.3 for which $0.3 < y_d < 0.9$ \citep[][]{Tea17, Rho18}.  
However, three galaxies (REBELS-12, 14, 39) require $y_d > 1 M_\odot$, which is likely inconsistent with pure SN production, and might require dust growth via accretion of heavy elements from the ISM \citep[e.g.][]{Mancini15, Popping17, Graziani20}. 
%
% TABLE MILKY WAY RESULTS
%
\begin{table*}
\begin{minipage}{170mm}
\begin{center}
\caption{Best-fit measured and derived REBELS galaxy properties for a MW extinction curve and different geometries. }
%\begin{tabular}{rrrccrrcccccc}
%\hline\hline
%\multicolumn{13}{c}{\code{SLAB GEOMETRY}}\\
%\multicolumn{4}{c}{\textit{Measured}}& \multicolumn{1}{c}{ID\#} & \multicolumn{8}{c}{\textit{Derived}}\\
%\cline{1-4} \cline{6-13} 
%$\beta$& $F_{1500}$& $F_{158}$& $M_*$    &  &$\tau_{1500}$ & $T_{1500}$ & SFR& $L_{\rm IR}$& $T'_d$& $M'_d$& $r_d$& $y_d$\\
%\hline
%       & $\mu$Jy   & $\mu$Jy  & $M_\odot$&  &  &  & $M_\odot$ yr$^{-1}$& $L_\odot$ &  K  &$M_\odot$& kpc&  $M_\odot$ \\
%\hline
\begin{tabular}{rrrccrrccccccc}
\hline\hline
\multicolumn{14}{c}{\code{SLAB GEOMETRY}}\\
\multicolumn{4}{c}{\textit{Measured}}& \multicolumn{1}{c}{ID\#} & \multicolumn{9}{c}{\textit{Derived}}\\
\cline{1-4} \cline{6-14} 
$\beta$& $F_{1500}$& $F_{158}$& $M_*$    &  &$\tau_{1500}$ & $T_{1500}$ & SFR& $L_{\rm IR}$& $T'_d$& $M'_d$& $r_d$& $y_d$& $\kappa_s$\\
\hline
       & $\mu$Jy   & $\mu$Jy  & $M_\odot$&  &  &  & $M_\odot$ yr$^{-1}$& $L_\odot$ &  K  &$M_\odot$& kpc&  $M_\odot$&  \\
\hline
 $-$1.29&  0.315&   67.2&  1.45e+09&  REBELS-05&  7.755&  0.109&   81.9& 8.57e+11&   60.9& 5.91e+06&   0.174&     0.22& 10.1\\
 $-$1.24&  0.329&   76.7&  3.16e+09&  REBELS-06&  8.438&  0.095&  105.2& 1.12e+12&   61.8& 7.05e+06&   0.190&     0.12& 10.8\\
 $-$2.17&  0.363&  101.4&  1.05e+09&  REBELS-08&   ... &   ... &   ... &      ...&   ... &     ... &    ... &      ...& ...\\
 $-$1.99&  0.543&   86.8&  8.71e+08&  REBELS-12&  2.085&  0.488&   37.7& 2.27e+11&   36.7& 3.47e+07&   0.508&     2.11& 0.9\\
 $-$2.21&  0.704&   60.0&  5.37e+08&  REBELS-14&  0.924&  0.719&   31.5& 1.04e+11&   32.6& 3.38e+07&   0.693&     3.33& 1.0\\
 $-$1.34&  0.448&   52.9&  3.09e+09&  REBELS-18&  7.143&  0.125&  129.5& 1.33e+12&   67.3& 5.01e+06&   0.160&     0.09& 18.8\\
 $-$2.33&  0.242&   71.2&  6.17e+08&  REBELS-19&   ... &   ... &   ... &      ...&   ... &     ... &    ... &      ...& ...\\
 $-$1.85&  0.263&  259.5&  7.76e+09&  REBELS-25&   ... &   ... &   ... &      ...&   ... &     ... &    ... &      ...& ...\\
 $-$1.79&  0.359&   50.6&  4.90e+09&  REBELS-27&  3.301&  0.335&   34.4& 2.69e+11&   46.5& 9.45e+06&   0.235&     0.10& 2.8\\
 $-$1.61&  0.547&   56.1&  4.17e+09&  REBELS-29&  4.591&  0.233&   69.4& 6.25e+11&   58.1& 5.68e+06&   0.174&     0.07& 9.0\\
 $-$1.50&  0.313&   60.4&  3.55e+09&  REBELS-32&  5.514&  0.183&   50.9& 4.88e+11&   53.4& 7.42e+06&   0.197&     0.11& 5.0\\
 $-$2.18&  0.404&  163.0&  3.80e+09&  REBELS-38&   ... &   ... &   ... &      ...&   ... &     ... &    ... &      ...& ...\\
 $-$1.96&  0.797&   79.7&  3.63e+08&  REBELS-39&  2.256&  0.462&   52.9& 3.34e+11&   44.2& 1.59e+07&   0.335&     2.31& 2.8\\
 $-$1.44&  0.302&   48.3&  3.02e+09&  REBELS-40&  6.077&  0.160&   64.4& 6.35e+11&   58.2& 5.73e+06&   0.172&     0.10& 8.2\\
\hline\hline
\multicolumn{13}{c}{\code{SPHERICAL GEOMETRY}}\\
\multicolumn{4}{c}{\textit{Measured}}& \multicolumn{1}{c}{ID\#} & \multicolumn{9}{c}{\textit{Derived}}\\
\cline{1-4} \cline{6-14} 
$\beta$& $F_{1500}$& $F_{158}$& $M_*$    &  &$\tau_{1500}$ & $T_{1500}$ & SFR& $L_{\rm IR}$& $T'_d$& $M'_d$& $r_d$& $y_d$& $\kappa_s$\\
\hline
       & $\mu$Jy   & $\mu$Jy  & $M_\odot$&  &  &  & $M_\odot$ yr$^{-1}$& $L_\odot$ &  K  &$M_\odot$& kpc&  $M_\odot$&  \\
\hline
 $-$1.29&  0.315&   67.2& 1.45e+09&  REBELS-05&  3.467&  0.099&   90.4& 9.56e+11&   62.5& 5.61e+06&   0.165&     0.21& 11.4\\
 $-$1.24&  0.329&   76.7& 3.16e+09&  REBELS-06&  3.637&  0.088&  113.0& 1.21e+12&   62.9& 6.80e+06&   0.180&     0.11& 11.7\\
 $-$2.17&  0.363&  101.4& 1.05e+09&  REBELS-08&   ... &   ... &   ... &      ...&   ... &     ... &    ... &      ...& ...\\
 $-$1.99&  0.543&   86.8& 8.71e+08&  REBELS-12&  1.186&  0.463&   39.7& 2.51e+11&   37.9& 3.12e+07&   0.547&     1.89& 1.2\\
 $-$2.21&  0.704&   60.0& 5.37e+08&  REBELS-14&  0.539&  0.709&   31.9& 1.09e+11&   33.2&	4.72E+07&	0.788&	   3.12& 1.2\\
 $-$1.34&  0.448&   52.9& 3.09e+09&  REBELS-18&  3.298&  0.111&  145.4& 1.52e+12&   69.4& 4.72e+06&   0.152&     0.08& 22.0\\
 $-$2.33&  0.242&   71.2& 6.17e+08&  REBELS-19&   ... &   ... &   ... &      ...&   ... &     ... &    ... &      ...& ...\\
 $-$1.85&  0.263&  259.5& 7.76e+09&  REBELS-25&   ... &   ... &   ... &      ...&   ... &     ... &    ... &      ...& ...\\
 $-$1.79&  0.359&   50.6& 4.90e+09&  REBELS-27&  1.810&  0.304&   37.9& 3.10e+11&   48.3& 8.58e+06&   0.243&     0.09& 3.6\\
 $-$1.61&  0.547&   56.1& 4.17e+09&  REBELS-29&  2.395&  0.205&   78.9& 7.36e+11&   60.5& 5.23e+06&   0.173&     0.07& 11.5\\
 $-$1.50&  0.313&   60.4& 3.55e+09&  REBELS-32&  2.760&  0.160&   58.3& 5.75e+11&   55.7& 6.78e+06&   0.190&     0.10& 6.3\\
 $-$2.18&  0.404&  163.0& 3.80e+09&  REBELS-38&   ... &   ... &   ... &      ...&   ... &     ... &    ... &      ...& ...\\
 $-$1.96&  0.797&   79.7& 3.63e+08&  REBELS-39&  1.278&  0.435&   56.1& 3.72e+11&   45.6& 1.47e+07&   0.364&     2.14& 3.5\\
 $-$1.44&  0.302&   48.3& 3.02e+09&  REBELS-40&  2.961&  0.140&   73.6& 7.43e+11&   60.5& 5.28e+06&   0.165&     0.09& 10.1\\
\hline
\label{tab:MW}
\end{tabular}
\end{center}
\end{minipage}
\tablecomments{$F_{158}$ fluxes are uncorrected for CMB effects. $L_{\rm IR}$ is computed from UV properties using eq. \ref{eq:L_IR}; it therefore might differ from the value obtained by Inami et al. 2021, in prep. which uses a different method.}
\end{table*}

\subsubsection{Spherical geometry}
Let us now consider the spherical case. The key difference between the two cases is due to the fact that, for a given extinction curve, spherical geometries are more opaque (see Fig. \ref{Fig:01}). This implies that a given $\beta$ value 
can be reproduced with a smaller $\tau_{1500}$, and yields a slighly lower $T_{1500}$ (see Tab. \ref{tab:MW}). The resulting star formation rates are essentially unaltered, and now span the range $31.9 < {\rm SFR}/(\msunyr) < 145.4$; REBELS-14 (REBELS-18) are confirmed to be the least (most) star forming system. Quantities related to dust differ only by a few percent; specifically, we find on average fractional differences of  ($12, 3, 8$)\% for $(L_{\rm IR}, T'_d, M'_d)$, respectively. Similar small differences are found also for $r_d$ and $y_d$. We can conclude that for a MW extinction curve, our results are not particularly sensitive to RT effects related to dust geometry. We confirm that no solutions can be found for the same 4 galaxies as in the slab case.    

%
% FIGURE 4
%
\begin{figure*}
\centering\includegraphics[width=0.9\textwidth]{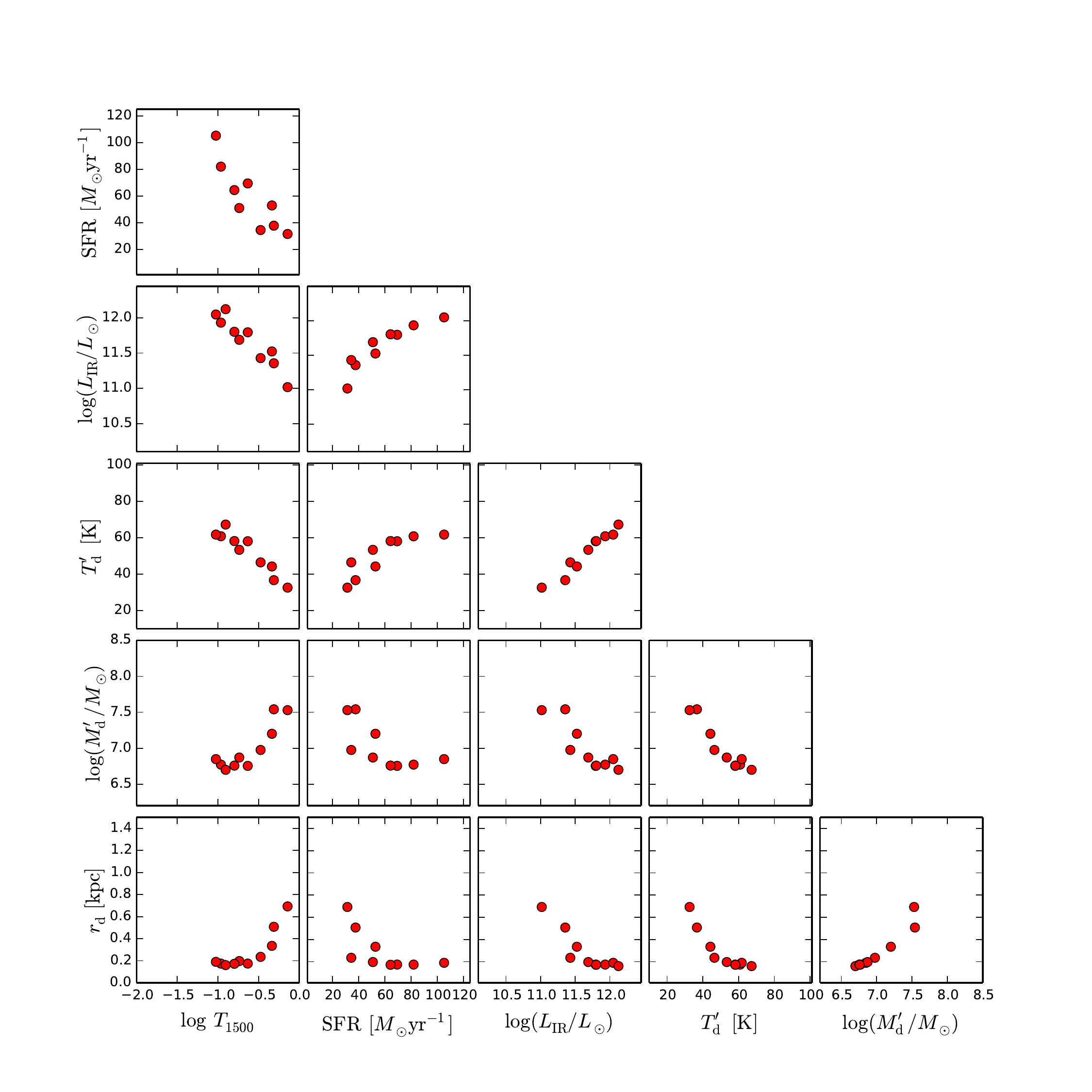}
\caption{Overview of the model results for the MW extinction curve and slab geometry case. The shown properties are those reported in Tab. \ref{tab:MW} for the 10 REBELS galaxies (red points) for which a solution can be found.}
\label{Fig:04}
\end{figure*}

\subsection{SMC extinction curve}
Adopting a SMC curve produces noticeable changes in the estimated galaxy properties reported in Tab. \ref{tab:SMC}. As it is clear from Fig. \ref{Fig:01}, the MW curve is more transparent than SMC, i.e. it requires a larger $\tau_{1500}$ to produce the same $\beta$ value. 
This feature is reflected in both the slab and spherical solutions which we discuss next; it also limits to 7 (3) the number of galaxies for which a self-consistent solution can be found in the slab (spherical) geometry cases explored.

\subsubsection{Slab geometry}
With respect to the MW, SMC extinction curves result in a lower UV optical depth ($\tau_{1500} < 2.7$) and larger transmissivity, or observed star formation fraction (eq. \ref{eq:fobs}), $T_{1500} > 0.42$. In turn, these properties determine SFR values that are on average $\approx 3$ times lower, with the largest discrepancy (factor $\approx 4$) found for the most absorbed galaxies (such as REBELS-06, $\tau_{1500}=2.71$), or equivalently reddest UV slope. 

The total IR luminosity/SFR ratio, is on average lower than for MW curves ($9.6 \simlt \log\, {\cal K}_{\rm IR} \simlt 9.8$); it similarly shows an increasing trend with SFR, however without reaching the plateau observed for the MW case at high SFR.  Although surprising to a first sight, as SFRs are lower, this result is determined by the even more pronounced decrease of $L_{\rm IR}$ (compare the MW and SMC cases in Tab. \ref{tab:MW} and \ref{tab:SMC}) when SMC curves are adopted. As the galaxies are now less star-forming, dust temperatures drop considerably (sample-averaged $T'_d=37\pm 3$ K), while the mean dust mass is essentially unchanged, $M'_d=(1.6\pm 0.5)\times 10^7 M_\odot$. 
No major differences are found in terms of the dust radius and SN yield. We confirm a compact, but slightly more extended configuration in which $r_d < 0.69$ kpc (REBELS-27 having the largest radius) in all galaxies. We find $0.14 \le y_d/M_\odot \le 0.57$, with mean value $\langle y_d \rangle = 0.29 M_\odot$, i.e. about a factor 3 below the MW value. In this case the need for dust growth in the ISM is sensibly decreased.

\subsubsection{Spherical geometry}
The differences between the slab and spherical geometries are minor. Although in this geometry galaxies are more transparent, as inferred by comparing $\tau_{1500}$ values in Tab. \ref{tab:SMC}, all other quantities are virtually the same as in the slab case, with gaps amounting to a few percent. These are even smaller than for MW, for which the larger opacities amplify the differences induced by geometry. This configuration allows solutions only for 3 galaxies, and therefore its statistical significance is rather weak.  

%
% TABLE SMC RESULTS
%
\begin{table*}
\begin{minipage}{170mm}
\begin{center}
\caption{Measured and model-predicted REBELS galaxy properties for a SMC extinction curve and different geometries.}
%\begin{tabular}{rrrccrrcccccc}
%\hline\hline
%\multicolumn{13}{c}{\code{SLAB GEOMETRY}}\\
%\multicolumn{4}{c}{\textit{Measured}}& \multicolumn{1}{c}{ID\#} & \multicolumn{8}{c}{\textit{Derived}}\\
%\cline{1-4} \cline{6-13} 
%$\beta$& $F_{1500}$& $F_{158}$& $M_*$    &  &$\tau_{1500}$ & $T_{1500}$ & SFR& $L_{\rm IR}$& $T'_d$& $M'_d$& $r_d$& $y_d$\\
%\hline
%       & $\mu$Jy   & $\mu$Jy  & $M_\odot$&  &  &  & $M_\odot$ yr$^{-1}$& $L_\odot$ &  K  &$M_\odot$& kpc&  $M_\odot$ \\
%\hline
\begin{tabular}{rrrccrrccccccc}
\hline\hline
\multicolumn{14}{c}{\code{SLAB GEOMETRY}}\\
\multicolumn{4}{c}{\textit{Measured}}& \multicolumn{1}{c}{ID\#} & \multicolumn{9}{c}{\textit{Derived}}\\
\cline{1-4} \cline{6-14} 
$\beta$& $F_{1500}$& $F_{158}$& $M_*$    &  &$\tau_{1500}$ & $T_{1500}$ & SFR& $L_{\rm IR}$& $T'_d$& $M'_d$& $r_d$& $y_d$& $\kappa_s$\\
\hline
       & $\mu$Jy   & $\mu$Jy  & $M_\odot$&  &  &  & $M_\odot$ yr$^{-1}$& $L_\odot$ &  K  &$M_\odot$& kpc&  $M_\odot$&  \\
\hline

 $-$1.29&  0.315&   67.2& 1.45e+09&  REBELS-05&  2.586&  0.436&   20.5& 1.36e+11&   36.9& 1.55e+07&   0.445&     0.57& 1.4\\
 $-$1.24&  0.329&   76.7& 3.16e+09&  REBELS-06&  2.712&  0.420&   23.7& 1.61e+11&   36.4& 2.04e+07&   0.504&     0.34& 1.2\\
 $-$2.17&  0.363&  101.4& 1.05e+09&  REBELS-08&   ... &   ... &   ... &      ...&	  ... &     ... &    ... &     ... & ...\\
 $-$1.99&  0.543&   86.8& 8.71e+08&  REBELS-12&   ... &   ... &   ... &      ...&	  ... &     ... &    ... &     ... & ...\\
 $-$2.21&  0.704&   60.0& 5.37e+08&  REBELS-14&   ... &   ... &   ... &      ...&	  ... &     ... &    ... &     ... & ...\\
 $-$1.34&  0.448&   52.9& 3.09e+09&  REBELS-18&  2.461&  0.453&   35.6& 2.29e+11&   42.3& 1.14e+07&   0.387&     0.20& 3.3\\
 $-$2.33&  0.242&   71.2& 6.17e+08&  REBELS-19&   ... &   ... &   ... &      ...&   ... &     ... &    ... &     ... & ...\\
 $-$1.85&  0.263&  259.5& 7.76e+09&  REBELS-25&   ... &   ... &   ... &      ...&	  ... &     ... &    ... &     ... & ...\\
 $-$1.79&  0.359&   50.6& 4.90e+09&  REBELS-27&  1.379&  0.632&   18.2& 7.88e+10&   31.8& 2.44e+07&   0.687&     0.26& 0.9\\
 $-$1.61&  0.547&   56.1& 4.17e+09&  REBELS-29&  1.803&  0.553&   29.2& 1.53e+11&   39.9& 1.09e+07&   0.416&     0.14& 3.0\\
 $-$1.50&  0.313&   60.4& 3.55e+09&  REBELS-32&  2.068&  0.510&   18.3& 1.05e+11&   34.4& 1.89e+07&   0.524&     0.28& 1.0\\
 $-$2.18&  0.404&  163.0& 3.80e+09&  REBELS-38&   ... &   ... &   ... &      ...&	  ... &     ... &    ... &     ... & ... \\
 $-$1.96&  0.797&   79.7& 3.63e+08&  REBELS-39&   ... &   ... &   ... &      ...&	  ... &     ... &    ... &     ... & ... \\
 $-$1.44&  0.302&   48.3& 3.02e+09&  REBELS-40&  2.215&  0.488&   21.1& 1.27e+11&   37.1& 1.44e+07&   0.447&     0.25& 1.6\\
\hline\hline
\multicolumn{13}{c}{\code{SPHERICAL GEOMETRY}}\\
\multicolumn{4}{c}{\textit{Measured}}& \multicolumn{1}{c}{ID\#} & \multicolumn{8}{c}{\textit{Derived}}\\
\cline{1-4} \cline{6-14} 
$\beta$& $F_{1500}$& $F_{158}$& $M_*$    &  &$\tau_{1500}$ & $T_{1500}$ & SFR& $L_{\rm IR}$& $T'_d$& $M'_d$& $r_d$& $y_d$& $\kappa_s$\\
\hline
       & $\mu$Jy   & $\mu$Jy  & $M_\odot$&  &  &  & $M_\odot$ yr$^{-1}$& $L_\odot$ &  K  &$M_\odot$& kpc&  $M_\odot$ &\\
\hline
 $-$1.29&  0.315&   67.2& 1.45e+09&  REBELS-05&  1.324&  0.437&   20.5& 1.35e+11&   36.9& 1.56e+07&   0.512&     0.57& 1.5\\
 $-$1.24&  0.329&   76.7& 3.16e+09&  REBELS-06&  1.381&  0.421&   23.7& 1.61e+11&   36.3& 2.05e+07&   0.578&     0.34& 1.3\\
 $-$2.17&  0.363&  101.4& 1.05e+09&  REBELS-08&   ... &   ... &   ... &      ...&	  ... &     ... &    ... &     ... & ...\\
 $-$1.99&  0.543&   86.8& 8.71e+08&  REBELS-12&   ... &   ... &   ... &      ...&	  ... &     ... &    ... &     ... & ...\\
 $-$2.21&  0.704&   60.0& 5.37e+08&  REBELS-14&  	... &   ... &   ... &      ...&   ... &     ... &    ... &     ... & ...\\
 $-$1.34&  0.448&   52.9& 3.09e+09&  REBELS-18&  	... &   ... &   ... &      ...&   ... &     ... &    ... &     ... & ...\\
 $-$2.33&  0.242&   71.2& 6.17e+08&  REBELS-19&   ... &   ... &   ... &      ...&	  ... &     ... &    ... &     ... & ...\\
 $-$1.85&  0.263&  259.5& 7.76e+09&  REBELS-25&   ... &   ... &   ... &      ...&	  ... &     ... &    ... &     ... & ...\\
 $-$1.79&  0.359&   50.6& 4.90e+09&  REBELS-27&  	... &   ... &   ... &      ...&   ... &     ... &    ... &     ... & ...\\
 $-$1.61&  0.547&   56.1& 4.17e+09&  REBELS-29&  0.957&  0.554&   29.2& 1.53e+11&   39.8& 1.09e+07&   0.491&     0.14&  3.5\\
 $-$1.50&  0.313&   60.4& 3.55e+09&  REBELS-32&   ... &   ... &   ... &      ...&	  ... &     ... &    ... &     ... & ...\\
 $-$2.18&  0.404&  163.0& 3.80e+09&  REBELS-38&   ... &   ... &   ... &      ...&	  ... &     ... &    ... &     ... & ...\\
 $-$1.96&  0.797&   79.7& 3.63e+08&  REBELS-39&  	... &   ... &   ... &      ...&   ... &     ... &    ... &     ... & ...\\
 $-$1.44&  0.302&   48.3& 3.02e+09&  REBELS-40&  	... &   ... &   ... &      ...&   ... &     ... &    ... &     ... & ...\\
\hline
\label{tab:SMC}
\end{tabular}
\end{center}
\end{minipage}
\tablecomments{$F_{158}$ fluxes are uncorrected for CMB effects. $L_{\rm IR}$ is computed from UV properties using eq. \ref{eq:L_IR}; it therefore might differ from the value obtained by Inami et al. 2021, in prep. which uses a different method.}
\end{table*}

\begin{figure*}
\centering
\includegraphics[width=0.49\textwidth]{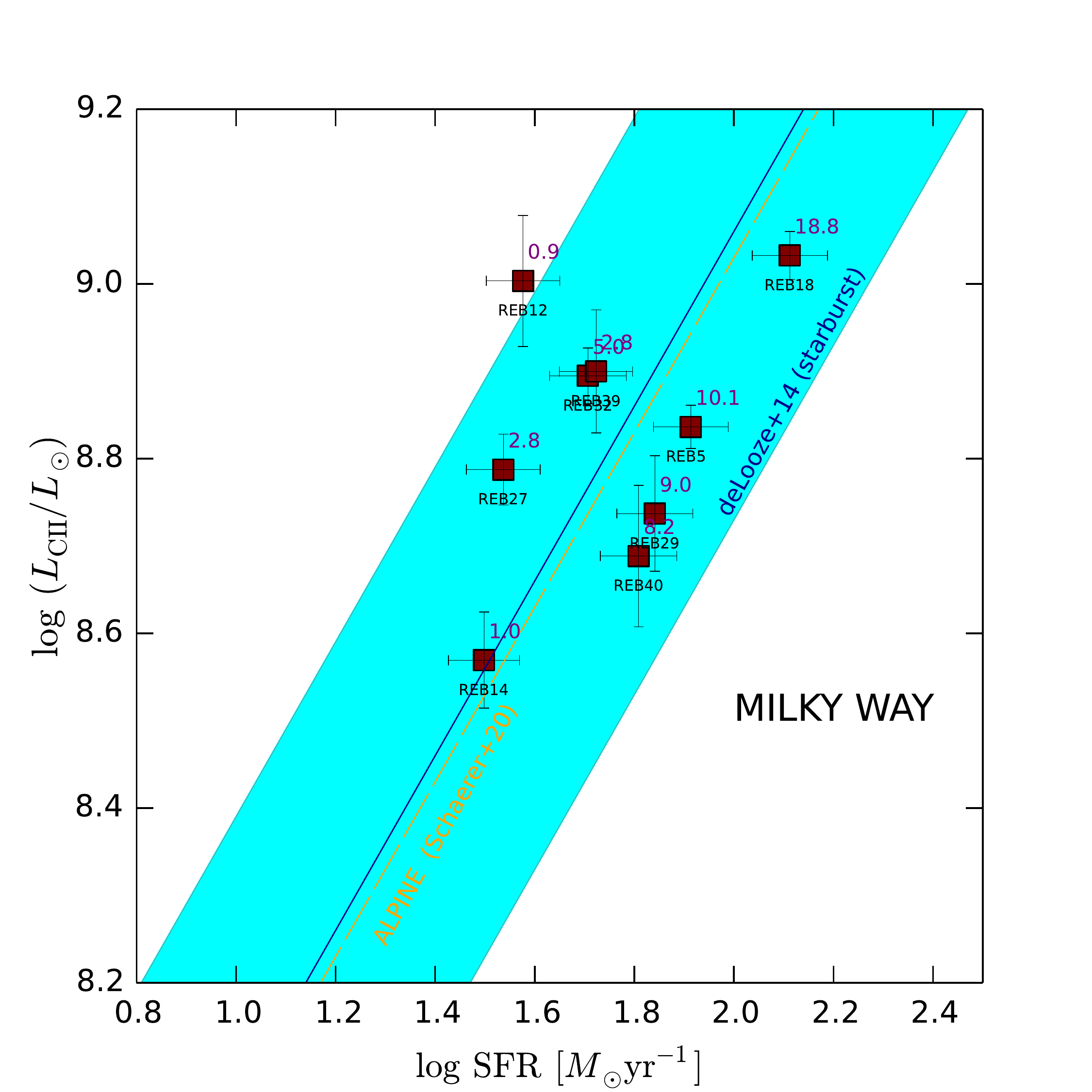}
\includegraphics[width=0.49\textwidth]{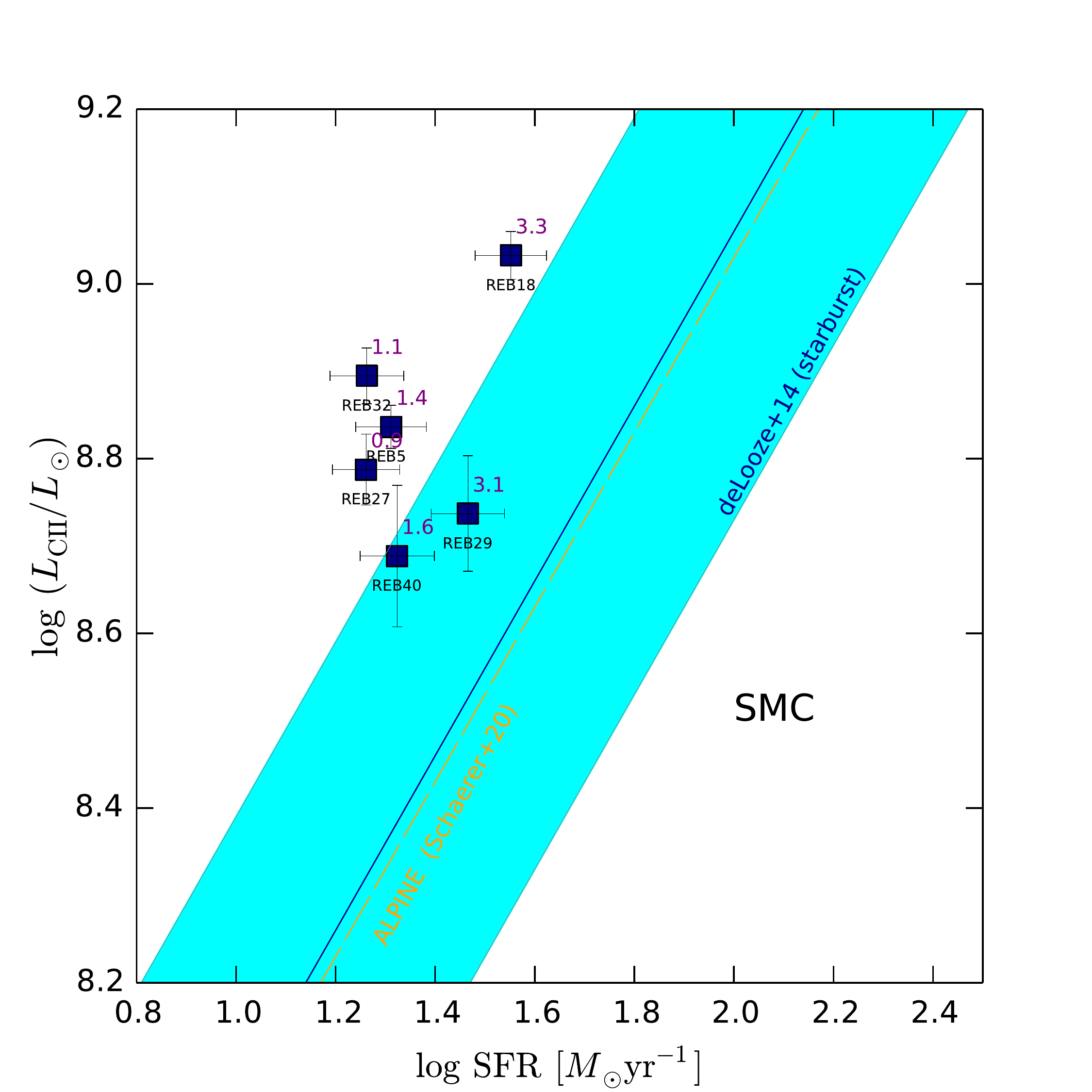}
\caption{\textit{Left panel:} [CII] luminosity as a function of the predicted SFR for 9 REBELS galaxies (red points) for a MW extinction curve and slab geometry; the $L_{\rm CII}$ values and errors are measured by the REBELS survey.  Among the 10 galaxies with a solution, REBELS-06 is undetected in [CII] and therefore does not appear in the plot. The REBELS galaxies ($z\approx 7$) closely follow the local starburst relation \citep[][aqua band]{deLooze14}. Also shown for completeness is the ALPINE \citep[][orange dashed]{Schaerer20} relation for $4.4 < z < 5.9$ galaxies. For each galaxy we give the corresponding burstiness parameter, $\ks$, value (purple) quantifying the upward deviation from the Kennicutt-Schmidt relation (eq. \ref{eq:ks} with $y_P=2$) \textit{Right:} Same as left panel, for a SMC curve and slab geometry. In this case the REBELS galaxies appear to be off and above the relation.}
\label{Fig:05}
\end{figure*}

\subsection{General trends}\label{sec:trends}
To conclude the analysis of the model, we discuss the general trends among the various derived physical quantities. As they are similar in the MW and SMC cases, we quantitatively refer for brevity only to the MW one. The corner plot in Fig. \ref{Fig:04} provides a bird's eye view of the trends discussed here. 
A few noticeable relations are worth highlighting. 

The transparency of galaxies, quantified by $T_{1500}$, increases with $r_{\rm d}$ and dust mass, but it decreases with dust temperature, and SFR. As in compact configurations the dust column is larger, more UV gets absorbed (decreasing $T_{1500}$), leading to more efficient dust heating, and hence higher dust temperatures.

While $M'_{d}$ does not depend strongly on SFR, actively star-forming systems have higher dust temperatures, and consequently $L_{\rm IR}$. In fact, $L_{\rm IR}$ seems to correlate very well with $T'_{\rm d}$ and anti-correlate with dust mass. In turn, higher dust temperatures are found in more compact, less dusty systems for which the SFR per unit area is larger. 

To summarise we can broadly divide the galaxies in the sample (at least those for which our model yields a solution; for a discussion on the remaining systems see Sec. \ref{sec:failure}) in two classes. In reality, the galaxy properties fall along a continuum sequence (see Fig. \ref{Fig:04}); however, the differences at the extremes justify the introduction of such rough separation. 

The first class (Class I) contains galaxies that are compact, and have large SFR (hence, particularly high $\Sigma_{\rm SFR} \simgt 300\, M_\odot {\rm kpc}^{-2}$), and low $I_m$ values. They are opaque, and have large IR luminosities; their dust is warm but present in limited amounts. The prototypical example of Class I is REBELS-18. The second class (Class II) contains systems that are more extended, moderately star forming and transparent, have a low $\Sigma_{\rm SFR} \simlt 300M_\odot {\rm kpc}^{-2}$ and large $I_m$ values. As a result their $L_{\rm IR}$ is about 4 times smaller than the corresponding Class I objects. Dust in these galaxies is cooler but more abundant. A representative object of Class II is REBELS-14. Whether such scenario corresponds to an evolutionary sequence remains to be ascertained. A precise determination of the stellar mass and age is therefore crucial to either support or discard this hypothesis.

\section{Additional implications}\label{sec:implications}
In the following we explore additional implications of our results by using empirical data not included in the model so far. These are the [CII]  158$\mu$m line-SFR relation, and the deviation from the Kennicutt-Schmidt relation, i.e. the burstiness of the REBELS galaxies. 

\subsection{[CII] line luminosity -- SFR relation}
For many targets, along with the $\approx 158\mu$m dust continuum, the REBELS survey has also measured the [CII]  158$\mu$m line flux. There are indications that high-$z$ galaxies align on a well-defined relation between the line luminosity, $L_{\rm CII}$, and SFR \citep[e.g.][]{Carniani18,Carniani20}. This relation was first discovered locally \citep{deLooze14}, but it has recently firmly confirmed for 118 galaxies at $4.4 < z < 5.9$ by the ALPINE collaboration  \citep{Schaerer20}. 
The original fit for starburst galaxies suggested by \citet[][]{deLooze14} is given by
\begin{equation}\label{eq:delooze}
\log\, L_{\rm CII} = 7.06 \pm 0.33 + (1.0 \pm 0.04) \log\, \rm SFR;
\end{equation}
here we adopt this relation, but note that \citet[][]{Carniani20} find a slightly larger zero-point 1$\sigma$ dispersion of $\pm 0.48$ dex. According to these authors, such larger dispersion may be associated with the presence of kpc-scale sub-components that are not common in the local Universe.

For completeness we show in Fig. \ref{Fig:05} also the ALPINE relation given in \citet[][Tab. A1]{Schaerer20}:
\begin{equation}\label{eq:alpine}
\log\, L_{\rm CII} = 7.03 \pm 0.17 + (1.0 \pm 0.12) \log\, \rm SFR.
\end{equation}
We warn that such fit has been obtained with assumptions that are different from the ones adopted here: they use a SMC-like extinction curve and a fixed dust temperature $T_d=45$ K. Hence, the comparison with the ALPINE data is not fully consistent. 

As our model predicts the total SFR of a galaxy, we can use the measured $L_{\rm CII}$ to verify whether the REBELS galaxies follow the same trend also at their mean $z\approx 7$. Among the 14 galaxies listed in Tab. \ref{tab:DATA}, 13 have also a [CII] line measurement (REBELS-06 is undetected, see footnote \ref{foot:REB06}). We can then associate the predicted SFR to the measured $L_{\rm CII}$ for each of them and compare it with the relation in eq. \ref{eq:delooze}.
The relation is shown in Fig. \ref{Fig:05} along with the model (SFR) and data ($L_{\rm CII}$) points for the 13 REBELS galaxies. They are calculated for MW (Tab. \ref{tab:MW}) and SMC (Tab. \ref{tab:SMC}) extinction curves in the slab geometry case.

The 9 REBELS galaxies with a solution and a [CII] detection nicely follow the local starburst galaxies relation if a MW extinction curve is adopted. This is far from trivial because the model does not use the [CII] line information at all. Moreover, this result strengthens the basis of the new method \citep{Sommovigo21} to infer the dust temperature combining [CII] line and continuum luminosity. Such method is in fact based on the assumption that eq. \ref{eq:delooze} holds also in the EoR.

%For completeness we include galaxies for which no solution has been found in terms of their dust properties (open squares). We note that these objects lie well above the relation, perhaps signalling that a reliable SFR determination can only be obtained by simultaneously considering both the UV and IR constraints on dust properties. 

For a SMC curve, instead, all REBELS galaxies lie considerably off and above the relation. Note that in general high-$z$ galaxies tend to lie \textit{below} the relation\footnote{See discussion in \citet{Ferrara19} and \citet[][]{Carniani18}.}. We consider this as an indication that galaxies in the REBELS sample might preferentially have a MW-like extinction curve.  

\subsection{Burstiness \& gas depletion time}
Using the model results, we can also investigate whether REBELS galaxies are starbursting. To quantify this statement we follow \citet[][]{Ferrara19} and introduce the \quotes{burstiness} parameter $\ks$, accounting for upward deviations from the Kennicutt-Schmidt (KS) average relation\footnote{The star formation rate (gas mass) per unit area, $\Sigma_{\rm SFR}$, ($\Sigma_{g}$) is expressed in units of $ M_\odot~{\rm yr}^{-1}~{\rm kpc}^{-2}$ ($M_\odot~{\rm kpc}^{-2}$).} \citep[e.g.][]{Heiderman10}:
\be\label{eq:ks}
\Sigma_{\rm SFR}= 10^{-12} \ks \Sigma_g^m \quad\quad (m=1.4).
\ee
Galaxies with $\ks > 1$ show a larger SFR per unit area with respect to those located on the KS relation having the same value of $\Sigma_g$, i.e. they tend to be starburst. Values of up to $\ks = 100$ have been measured for sub-millimeter galaxies, e.g. \citet{Hodge15, Vallini21}.

To compute $\ks$ from eq. \ref{eq:ks} we need two additional quantities: (a) the radius of the stellar, $r_*$, and gas, $r_g$, distribution; (b) the gas mass, $M_g$. For the former, we assume $r_g = y_P r_* = y_P r_d$ where $y_P$, {  referred to as the Perito ratio after \citet[][]{Carniani18}}, is a factor $\simgt 1$. The previous equalities are inspired by the empirical evidence that in high-$z$ galaxies, the gas (traced by [CII] emission) is more extended than the dust/stellar emitting regions which instead show a similar size \citep{Fujimoto19, Fujimoto20, Carniani20, Ginolfi20}. These studies suggest that $y_P \approx 2$; we adopt this value here, but note that the $\ks$ values shown in Fig. \ref{Fig:05} can be easily scaled to other $y_P$ choices recalling that, from eq. \ref{eq:ks} and the relation $r_g = y_P r_*$, $\ks \propto y_P^{2m} = y_P^{2.8}$. We then use $r_*$ and $r_g$ to write $\Sigma_{\rm SFR} = {\rm SFR}/\pi r_*^2$, and $\Sigma_g = M_g/\pi r_g^2$. This leads us to the final step, i.e. the determination of the gas mass which is deduced from the predicted dust mass: $M_g = M'_d/D$, where $D$ is the dust-to-gas ratio\footnote{We assume that $D\propto Z$. Then for the MW $D_{\rm MW}=1/162$ \citep{Remy14}, $Z=0.004$ implies $D = 1/575$. For completeness, the value for the SMC is $D_{\rm SMC}=1/1408$.} for $Z=0.004$.

The derived $\ks$ values are shown in Fig. \ref{Fig:05} by the numbers close to each galaxy point, and reported in Tab. \ref{tab:MW} and Tab. \ref{tab:SMC} for all cases. For the MW we find $0.9 < \ks < 18.8$, with most actively star forming systems showing larger deviations from the KS relation (i.e. they have larger $\ks$ values). On this basis, we conclude that REBELS galaxies, in addition to the $L_{\rm CII}$-SFR relation, approximately follow the KS one as well, although a few appear to be relatively bursty (as e.g. REBELS-18, $\kappa_s = 18.8$). This conclusion holds also for the SMC extinction curve: in fact, we find $0.9 < \ks < 3.3$. 

Finally, we can compute the mean gas depletion timescale, $t_d$, for the sample,  
\begin{equation}\label{eq:td}
    t_{\rm dep} = \frac{\Sigma_g}{\Sigma_{\rm SFR}} = \frac{M'_d}{D\, {\rm SFR}} y_P^{-2} = 0.11\,y_P^{-2} {\rm Gyr},
\end{equation}
having assumed ${\rm SFR}=65.8\, \msunyr$ and $M'_d=1.3 \times 10^7 M_\odot$ as the mean values for the MW/slab configuration.
This result can be compared with the extrapolation of the $z < 4$ relation found by \citet[][]{Tacconi18} from the PHIBBS survey for galaxies located on the main sequence\footnote{Of course, there is no guarantee that REBELS galaxies are located on such relation. In addition, strictly speaking, the depletion time in the PHIBBS sample is computed for molecular, rather than total gas mass. By using this formula we implicitly assume that $M_{\rm H2} \approx M_g$, which for such high-$z$ galaxies should represent a reasonable approximation \citep[][]{Tacconi18}.}
\begin{equation}\label{eq:tdobs}
    \log (t_{\rm dep}/{\rm Gyr}) = 0.09 - 0.62 \log (1+z).
\end{equation}
When evaluated at the REBELS sample mean redshift, $\langle z_s \rangle = 7.01$, eq. \ref{eq:tdobs} gives $t_{\rm dep} = 0.34$ Gyr, formally requiring a Perito ratio $y_P =0.57$. As such value is somewhat in tension with [CII] observations indicating $y_P \approx 2$, this result might indicate that the extrapolation of the PHIBBS relation into the EoR results in an overestimate of the depletion time. Further work is necessary to clarify this issue in detail.

\begin{comment}
% TABLE MEURER RESULTS
%
\begin{table}
\begin{minipage}{80mm}
\caption{Model-predicted REBELS galaxy properties for a Meurer attenuation curve. Quantities depending on grain optical properties, such as $M'_d$, $T'_d, r_d$ or $y_d$, cannot be obtained in this case.}
\begin{center}
\begin{tabular}{ccccc}
\hline\hline
ID\#    &$\tau_{1500}$ & $T_{1500}$ & SFR                & $L_{\rm IR}$\\\hline
        &       &       & $M_\odot$ yr$^{-1}$& $L_\odot$    \\
\hline
%REB04&	1.284&	0.652&	20.7&	8.46E+10\\
REB05&	1.715&	0.568&	15.2&	7.68E+10\\
REB06&	1.807&	0.552&	18.0&	9.47E+10\\
REB08&	1.121&	0.687&	16.4&	6.03E+10\\
REB12&	0.839&	0.753&	24.8&	7.19E+10\\
REB14&	0.526&	0.836&	22.5&	4.33E+10\\
REB18&	1.624&	0.585&	22.1&	1.08E+11\\
REB19&	1.315&	0.645&	10.6&	4.44E+10\\
REB25&	3.151&	0.370&	28.1&	2.07E+11\\
REB27&	0.799&	0.763&	14.0&	3.91E+10\\
REB29&	1.129&	0.685&	23.6&	8.72E+10\\
REB32&	1.331&	0.642&	10.4&	4.39E+10\\
%REB37&	1.523&	0.604&	28.0&	1.30E+11\\
REB38&	1.390&	0.630&	20.0&	8.70E+10\\
REB39&	0.589&	0.818&	24.1&	5.14E+10\\
\hline
\label{tab:Meurer}
\end{tabular}
\end{center}
\end{minipage}
\end{table}
\end{comment}
%
%
\section{Where the IRX relation fails}\label{sec:failure}
%Two-phase medium, reconnect with Ferrara+2016. Recall that ${\rm IRX} = (1- T_{\rm UV})/T_{\rm UV} = T_{\rm UV}^{-1}-1$.
%Discussed in Faisst+2017 (from Ilse talk); connect with spatial offsets between continuum and UV (also seen in SERRA) 

%
% FIGURE 6
%
\begin{figure}
\centering\includegraphics[scale=0.7]{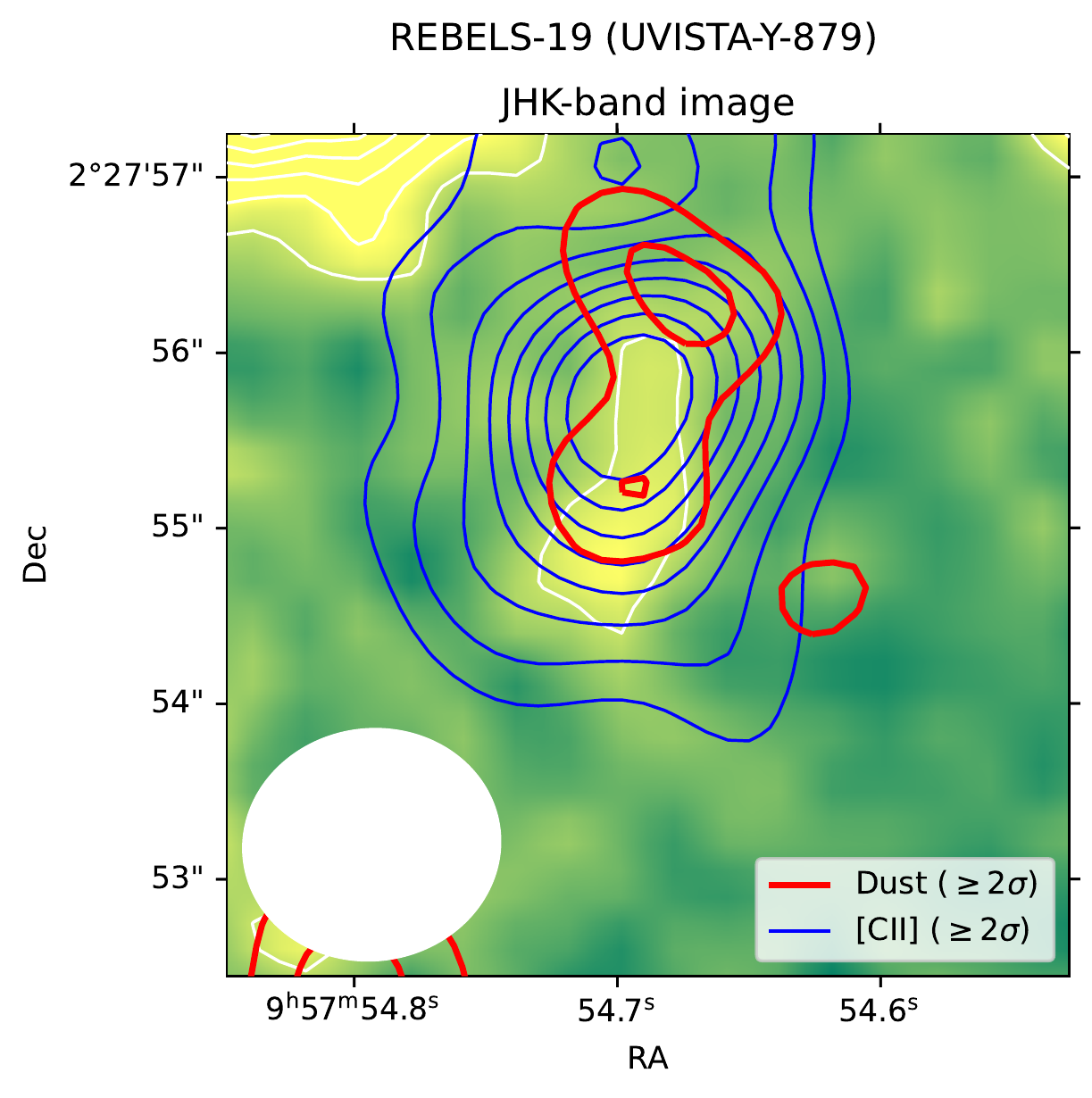}
\caption{FIR  158$\mu$m dust continuum (red) and [CII] line (blue) emission contours for REBELS-19 ($z=7.37$) superposed to the JHK-band composite image and (white) contours. The contours start from 2$\sigma$ and increase in steps of 1$\sigma$. The white ellipse at the bottom left shows the beam size; thin dotted line contours indicate negative sources. The spatial segregation between the FIR and the restframe UV emission is evident. For more details see Inami et al.  2021, in prep.}
\label{Fig:06}
\end{figure}

At the beginning of Sec. \ref{sec:Results} we have noted that a solution cannot be found with our method for some galaxies both for MW and SMC extinction curve cases, and we provided there a first qualitative explanation. Here we additionally note that such no-solution cases are characterized by large $I_m$ values. In other words, these peculiar galaxies have a very large IR-to-UV flux ratio compared to their UV slope. Let us see how this can be understood.

Using the depletion time definition in eq. \ref{eq:td} one can write an approximate expression for the SED \quotes{color}, i.e. the $F_{158}/F_{1500}$ ratio, predicted by the model by combining eq. \ref{eq:F158} and \ref{eq:SFR_obs}:
\begin{equation}\label{eq:SEDcolor}
    \frac{F_{158}}{F_{1500}} \simeq 7062 \frac{y}{T_{1500}} \exp\left[{-3\ \left(\frac{y}{1-T_{1500}}\right)^{1/6}}\right],
\end{equation}
where $y = (Z/Z_\odot)(t_{\rm dep}/\rm Gyr)$. Such expression is valid in the temperature range $T_{\rm CMB} \simlt T_d \simlt T_{158} = h_P\nu_{158}/k_B = 91.2$ K, as we have neglected the presence of the CMB in eq. \ref{eq:F158}. Eq. \ref{eq:SEDcolor} entails interesting physical implications. For a given $T_{1500}$ value, galaxies with a longer depletion time (recall that in this paper we have fixed $Z=0.28 Z_\odot$) have redder SED colors. Alternatively, for a given $t_{\rm dep}$, the ratio increases towards more opaque (lower $T_{1500}$) systems.

Let us now divide the above flux ratio by $(\beta - \beta_{\rm int}) = - \ln T_{1500} = \tau_{\rm eff}$, to obtain the quantity we have defined as the molecular index, $I_m$ in eq. \ref{eq:Im},
and take the optically thin limit, for which $1-T_{1500} \approx \tau_{\rm eff}$ and $T_{1500} \approx 1$. Then,
\begin{equation}\label{eq:Im2}
    I_{\rm m} = \frac{(F_{158}/F_{1500})}{(\beta-\beta_{\rm int})}\simeq 7062\ x e^{-3 x^{1/6}},
\end{equation}
where $x=y/\tau_{\rm eff}$. The previous expression has a maximum located at $x_{\rm max}=64$, corresponding to $I_m^* \equiv I_m(x_{\rm max}) \simeq 1120$ (for the MW). Hence, galaxies showing
values of $I_m > I_m^*$ cannot be reproduced by a single zone model. Physically, this depends on the fact that $F_{158}$ can be increased by raising the dust mass or the temperature. However, increasing $M'_d$ while keeping $(\beta - \beta_{\rm int})$ (i.e. the effective optical depth) low\footnote{Recall that $\beta_{\rm int}$ in our model is computed (Sec. \ref{sec:Method}) self-consistently with IMF, age, metallicity, $Z$ and ${\cal K}_{1500}$; therefore, it cannot be varied independently.} implies pushing the 
dust temperatures to values progressively closer to the CMB, thus preventing $\FIR$, and hence $I_m$, to increase beyond $I_m^*$. These findings are confirmed by the results reported in the Tables. Indeed, for the MW case, the no-solution galaxies all have $I_m > 1183$. 
%From the comparison between Tab. \ref{tab:DATA} and Tab. \ref{tab:SMC}, we find $I_m > 617$ for the SMC. 

We conclude that $I_m > I_m^*$ values can be achieved only if the FIR luminosity is spatially decoupled from the UV emitting regions. In such a scenario, the former is produced in optically thick, star forming clumps (likely, giant molecular complexes), and produce the high $F_{158}$ values required. The small UV optical depth is produced instead by the diffuse, interclump gas component in which young stars are embedded after they disperse their natal cloud. This two-phase configuration should then characterise no-solution galaxies.

We speculate that no-solution galaxies are extreme Class II systems, as defined in Sec. \ref{sec:trends},  which have developed a prominent two-phase ISM structure with spatially segregated IR and UV emitting regions throughout their disk. This hypothesis is supported by Fig. \ref{Fig:06}, showing the continuum and [CII] emission, along with the restframe UV image of REBELS-19, a no-solution case. The spatial displacement between the continuum and UV emission is clearly visible. Once observed at higher spatial resolution, other REBELS no-solution galaxies might show a similar spatially decoupled structure. As a final caveat, we note that the above scenario questions the use of the IRX-$\beta$ relation, which implicitly assumes that the IR and UV emission are co-spatial. According to our results such relation can be only safely applied to galaxies with $I_m \simlt I_m^*$ 

\section{Summary}\label{sec:summary}
We have analyzed the FIR dust continuum measurements obtained for 14 $z\approx 7$ galaxies by the ALMA REBELS Large Program, in combination with restframe UV data, with the aim of deriving the physical properties of these early systems. Our method uses as a input three measurements, i.e. (a) the UV spectral slope, $\beta$, (b) the observed UV continuum flux, $\FUV$ at $1500$\AA,  (c) the observed far-infrared continuum flux, $\FIR$, at $\approx 158\mu$m to derive 7 quantities ($\tau_{1500},\, T_{1500},\, {\rm SFR},\, L_{\rm IR},\, T'_d,\, M'_d,\, r_d$). An additional one, the dust yield per supernova, $y_d$, is obtained using external information on the galaxy stellar mass.

The results are summarized in Tab. \ref{tab:MW} and Tab. \ref{tab:SMC} for the MW and SMC curves, respectively, and for different dust/stellar relative geometries. They are also graphically shown in Fig. \ref{Fig:04}. 
In general, we find that different geometries have little impact on the parameter determination; changing the extinction curve has instead a significant impact. For example, the estimated SFR is $\approx 3\times$ lower for a SMC curve. We argue in Sec. \ref{sec:implications} that on the basis of the measured [CII] line luminosity, the MW extinction curve appears to be preferable. 

The key results for the fiducial (MW extinction case) are summarized as follows.
\begin{itemize}
\item[{\color{red} $\blacksquare$}] REBELS galaxies are \textit{physically} optically thick at 1500\AA\ but due to geometrical radiative transfer effects they are relatively transparent (i.e. low \textit{effective} optical depth, see Sec. \ref{subsec:beta-to-tau}), with $28-90.5$\% of the star formation being obscured\footnote{We prefer to define these galaxies as \quotes{relatively transparent} as (i) they are all detected in UV, and (b) because their V-band optical depth is $\simlt 3$, i.e. it is not extremely high as for obscured, sub-millimeter galaxies.}. The total star formation rates are in the range $31.5 < {\rm SFR}/(\msunyr) < 129.5$ with REBELS-14 (REBELS-18) being the least (most) star forming system.  

\item[{\color{red} $\blacksquare$}] The sample-averaged dust mass and temperature are $(1.3\pm 1.1)\times 10^7 M_\odot$ and $52 \pm 11$ K, respectively. However, in some galaxies dust is particularly abundant (REBELS-14, $M'_d \approx 3.4 \times 10^7 M_\odot$), or hot (REBELS-18, $T'_d \approx 67$ K). The dust distribution is compact with 70\% of the galaxies showing $r_d < 0.3$ kpc. Such compact configuration, in general, produces high dust temperatures: the hottest REBELS-18 indeed is the most compact system with $r_d = 0.16$ kpc. 

\item[{\color{red} $\blacksquare$}] By augmenting the method with stellar mass information obtained by the REBELS Collaboration from SED fitting, we estimate the dust yield per supernova, $y_d$. We find that $0.1 \le y_d/M_\odot \le 3.3$, with 70\% of the galaxies requiring $y_d < 0.25 M_\odot$. Three galaxies (REBELS-12, 14, 39) require $y_d > 1 M_\odot$, which is likely inconsistent with pure SN production, and might require dust growth via accretion of heavy elements from the ISM. { We warn that using non-parametric star formation histories might increase the stellar masses by $\approx 3$ times, thus reducing the above yields by the same factor.}

\item[{\color{red} $\blacksquare$}] With the SFR predicted by the model, REBELS galaxies detected in [CII] nicely follow the local   $L_{\rm CII}-$SFR relation \citep[][]{deLooze14} if a MW extinction curve is adopted. For a SMC curve, instead, all REBELS galaxies lie considerably off and above the relation. We also show that REBELS galaxies are approximately located on the Kennicutt-Schmidt relation (burstiness parameter $k_s \simlt 18.8$). The sample-averaged gas depletion time is of $0.11\, y_P^{-2}$ Gyr, where $y_P$ is the ratio of the gas-to-stellar distribution radius. 

\item[{\color{red} $\blacksquare$}] For some systems (4 in the case of the MW curve) a solution simultaneously matching the observed ($\beta, \FUV, \FIR$) values cannot be found. This occurs when the molecular index $I_m = F_{158}/\FUV(\beta-\beta_{\rm int})$ exceeds the threshold $I_m^* \approx 1120$ for a MW extinction curve. For these objects (REBELS-19 being the most outstanding example) we argue that the FIR luminosity is not co-spatial with the UV-emitting regions, questioning the use of the IRX-$\beta$ relation.  
\end{itemize}

\acknowledgments
AF acknowledges support from the ERC Advanced Grant INTERSTELLAR H2020/740120. Any dissemination of results must indicate that it reflects only the author’s view and that the Commission is not responsible for any use that may be made of the information it contains. Generous support from the Carl Friedrich von Siemens-Forschungspreis der Alexander von Humboldt-Stiftung Research Award is kindly acknowledged (AF). AF thanks the European Southern Observatory (ESO) and Max-Planck for Astrophysics (MPA) in Garching for a warm hospitality during part of this research.

MA acknowledges support from FONDECYT grant 1211951, ANID + PCI + INSTITUTO MAX PLANCK DE ASTRONOMIA MPG 190030, ANID + PCI + REDES 190194 and ANID BASAL project FB210003. 
HI and HSBA acknowledge support from the NAOJ ALMA Scientific Research Grant Code 2021-19A. HI acknowledges support from the JSPS KAKENHI Grant Number JP19K23462.
EdC gratefully acknowledges support from the Australian Research Council Centre of Excellence for All Sky Astrophysics in 3 Dimensions (ASTRO 3D), through project number CE170100013. PD acknowledges support from the European Research Council's starting grant ERC StG-717001 (``DELPHI"), from the NWO grant 016.VIDI.189.162 (``ODIN") and the European Commission's and University of Groningen's CO-FUND Rosalind Franklin program. RJB and MS acknowledge support from TOP grant TOP1.16.057. SS acknowledges support from the Nederlandse Onderzoekschool voor Astronomie (NOVA). IDL acknowledges support from ERC starting grant DustOrigin 851622.

This paper is based on data obtained with the ALMA Observatory, under the Large Program 2019.1.01634.L. ALMA is a partnership of ESO (representing its member states), NSF(USA) and NINS (Japan), together with NRC (Canada), MOST and ASIAA (Taiwan), and KASI (Republic of Korea), in cooperation with the Republic of Chile. The Joint ALMA Observatory is operated by ESO, AUI/NRAO and NAOJ.
All plots in this paper were built with the \textsc{matplotlib}  \citep{Hunter07} package for \textsc{PYTHON}. We gratefully acknowledge computational resources of the Center for High Performance Computing (CHPC) at SNS.  \\

{\bf Data Availability}\\
Data available on request.

\newpage
\bibliography{paper}{}
\bibliographystyle{aasjournal}

\appendix
\section{Dust temperature in optically thick media}\label{app_a}
\begin{figure}
\centering
\includegraphics[width=0.6\textwidth]{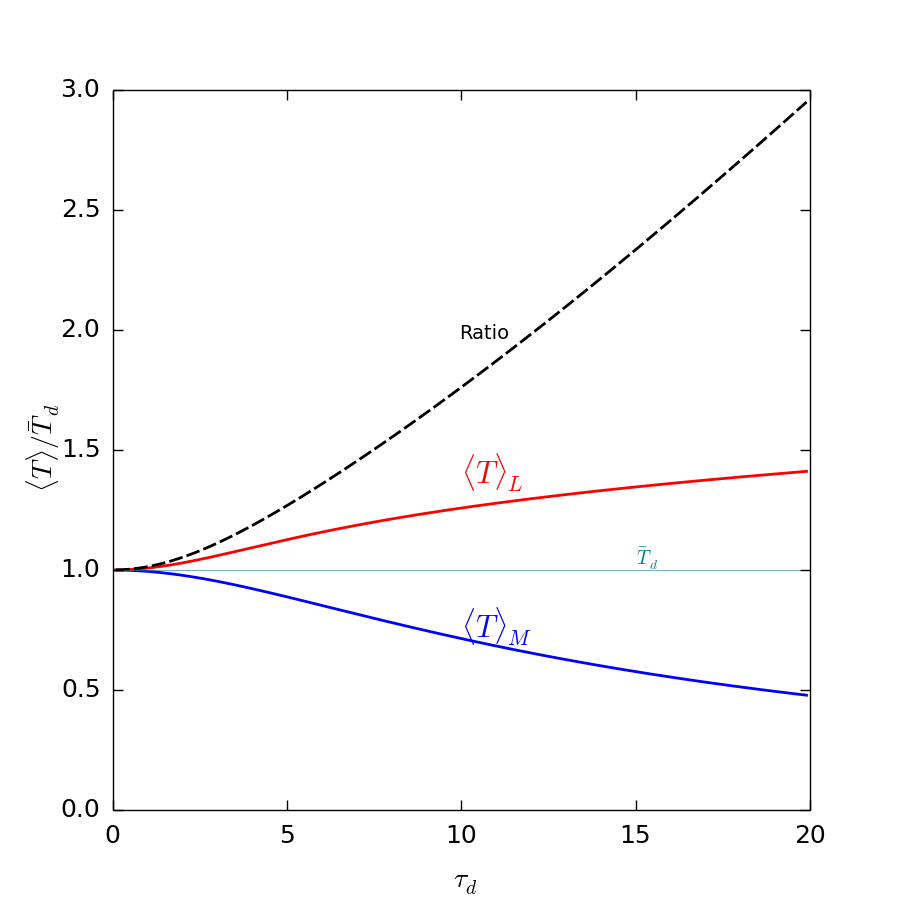}
\caption{Dependence of the luminosity- and mass-weighted temperatures (for $\beta_d=2$), along with their ratio, as a function of the total dust optical depth of the layer $\tau_d$ in units of the mean physical temperature, $\bar T_d$, defined in eq. \ref{eq:Tmpt}.}
\label{Fig:App_Fig01}
\end{figure}

Consider a uniform absorbing dust layer of total optical depth $\tau_d$, and divide it into $N$ equal slices, each with an optical depth $\Delta \tau=\tau_d/N$. The layer boundary ($N=0$) is illuminated by an external UV flux, $F_0$. The flux reaching the $i$-th slice is $F_i = F_{\rm 0} \exp(- \tau_i) = F_0 \exp(- i \Delta\tau)$; the absorbed flux is then $F_i(1-\exp(-\Delta\tau))$. Using eq. \ref{eq:Tmpt} for the mean physical temperature (MPT), and specializing for simplicity to the relevant case $\beta_d=2$, it is easy to show that the dust temperature (assuming that the entire slab is optically thin to re-emitted IR photons) in the $i$-th slice is
\begin{equation}\label{eq:Ti}
    T_i = \bar T_d \Bigg[N e^{-i \Delta\tau} \frac{(1-e^{-\Delta\tau})}{(1-e^{-\tau_d})}\Bigg]^{1/6}.
\end{equation}
The temperature therefore decreases with depth into the layer. We therefore define the \textit{mass-weighted} temperature by averaging\footnote{As the slices contain the same mass, this averaging is equivalent to a mass averaging} the non-constant factor over the depth $i$:
\begin{equation}
    \langle e^{-(\tau_d/6N)i}\rangle_M  = \frac{1}{N} \int_0^N e^{-(\tau_d/6N)i} di = \frac{6}{\tau_d} (1-e^{-\tau_d/6})
\end{equation}
Inserting the previous expression in eq. \ref{eq:Ti}, 
\begin{equation}
     \langle T_i \rangle_M = \bar T_d \Bigg[N \frac{(1-e^{-\tau_d/N})}{(1-e^{-\tau_d})}\Bigg]^{1/6} \frac{6}{\tau_d} (1-e^{-\tau_d/6}),
\end{equation}
further imposing that all slices are optically thin so that the MPT can be safely applied, $\Delta\tau \ll 1$ or, equivalently, $N\gg 1$, we find
\begin{equation}\label{eq:MeanTiM}
     {\lim_{N\rightarrow \infty}\, \langle T_i \rangle_M = \bar T_d \frac{6}{\tau_d^{5/6}} \frac{(1-e^{-\tau_d/6})}{(1-e^{-\tau_d})^{1/6}} \equiv \bar T_d f_M(\tau_d).}
\end{equation}
Note that if the layer is optically thin $\tau_d \rightarrow 0$, the above formula correctly returns $\langle T_i \rangle_M = \bar T_d$.  
Eq. \ref{eq:MeanTiM} is shown in Fig. \ref{Fig:App_Fig01}. We see that the MPT always overestimates the actual dust temperature, and it is correct only if the layer is optically thin. We note that the actual dust temperature
$\langle T_i \rangle$ can be substantially lower than $\bar T_d$: for example, if $\tau_d = 8$ (20), $\langle T_i \rangle_M = 0.78\, (0.48) \bar T_d$.

More often one wants to use the \textit{luminosity-weighted} temperature, i.e. the temperature of a single-temperature grey-body producing an IR luminosity equal to the absorbed UV luminosity input in the system. This is formally defined as 
\begin{equation}
     \langle T_i \rangle_L = \frac{\int_0^N T_i F_i di}{\int_0^N F_i di}.
\end{equation}
Recalling that $F_i \propto \Theta M_{d,i} T_i^6$, and taking again the limit $N\rightarrow \infty$, we find
\begin{equation}\label{eq:MeanTiL}
     \langle T_i \rangle_L = \bar T_d\, \frac{6}{7}\tau_d^{1/6}\, \frac{(1-e^{-7\tau_d/6})}{(1-e^{-\tau_d})^{7/6}} \equiv \bar T_d f_L(\tau_d).
\end{equation}
As for the mass-weighted temperature, $\langle T_i \rangle_L = \bar T_d$ for optically thin cells. However, the L-weighted temperature \textit{increases} as the cell becomes optically thick. For example, if $\tau_d = 8$ (20), $\langle T_i \rangle_L = 1.2\, (1.4)\, \bar T_d$. Hence, $\langle T_i \rangle_L$ is always larger than $\bar T_d$, the cell temperature one would derive by neglecting RT effects. Also, for optically thin cells $\bar T_d = \langle T_i \rangle_M = \langle T_i \rangle_L$; however as $\tau_d$ increases the mass- and luminosity-weighted temperatures diverge. The ratio between the two temperatures is given by $f_L/f_M$:
\begin{equation}\label{eq:ratio}
     {\cal R} = \frac{\langle T_i \rangle_L}{\langle T_i \rangle_M} = \frac{f_L}{f_M} =  \frac{\tau_d}{7}\, \frac{1}{(1-e^{-\tau_d})} \frac{(1-e^{-7\tau_d/6})}{(1-e^{-\tau_d})};
\end{equation}
For very optically thick cells, the ratio increases linearly, ${\cal R} \simeq \tau_d/7$. It is useful to note that the SED produced by a multi-temperature layer deviates from a single-$T$ graybody: the larger $\tau_d$, the larger is the deviation.  Such SED can be \textit{approximated} by a single-$T$ graybody  with temperature $T = \langle T_i \rangle_L$ provided that the dust mass is effectively decreased by the ratio $M_d (\bar T/\langle T_i \rangle_L)^6 = M_d f_L^{-6}$ so to produce the same total IR luminosity.

\end{document}

%% file: definitions.tex
\def\be{\begin{equation}}
\def\ee{\end{equation}}
\newcommand\code[1]{\textsc{\MakeLowercase{#1}}}

\newcommand\quotes[1]{``{#1}"}
\def\gsim{\lower.5ex\hbox{\gtsima}} 
\def\lsim{\lower.5ex\hbox{\ltsima}} 
\def\gtsima{$\; \buildrel > \over \sim \;$} 
\def\ltsima{$\; \buildrel < \over \sim \;$} \def\gsim{\lower.5ex\hbox{\gtsima}} 
\def\lsim{\lower.5ex\hbox{\ltsima}} 
\def\simgt{\lower.5ex\hbox{\gtsima}} 
\def\simlt{\lower.5ex\hbox{\ltsima}} 
\newcommand{\ped}[1]{_{\mathrm{#1}}}

\def\msunyr{\msun/{\rm yr}}

\def\mum{\mu {\rm m}}

\def\S*{$\Sigma_{\rm SFR}$}

\def\FUV{F_{\rm 1500}}
\def\FIR{F_{\rm 158}}

\def\msunyr{{\rm M}_\odot {\rm yr}^{-1}}

\def\ks{\kappa_{\rm s}}

\definecolor{apcolor}{HTML}{b3003b}
\definecolor{afcolor}{HTML}{800080}
\definecolor{lvcolor}{HTML}{DF7401}
\definecolor{mdcolor}{HTML}{01abdf} %mahsa and davide
\definecolor{cbcolor}{HTML}{ff0000}
\definecolor{sccolor}{HTML}{cc5500} %stefano
\definecolor{sgcolor}{HTML}{00cc7a}